\documentclass[twocolumn,amsmath,amssymb,floats,notitlepage]{revtex4-1}
\usepackage[latin9]{inputenc}
\usepackage{mathtools}
\usepackage{amsmath}
\usepackage{graphicx}
\usepackage{eufrak}
\usepackage[normalem]{ulem}
\usepackage{subfigure}
\makeatletter

\usepackage{comment}
\newcommand{\beq}{\begin{equation}}
\newcommand{\eeq}{\end{equation}}
\newcommand{\bea}{\begin{eqnarray}}
\newcommand{\eea}{\end{eqnarray}}

\@ifundefined{textcolor}{}{%
 \definecolor{BLACK}{gray}{0}
 \definecolor{WHITE}{gray}{1}
 \definecolor{RED}{rgb}{1,0,0}
 \definecolor{GREEN}{rgb}{0,1,0}
 \definecolor{BLUE}{rgb}{0,0,1}
 \definecolor{CYAN}{cmyk}{1,0,0,0}
 \definecolor{MAGENTA}{cmyk}{0,1,0,0}
 \definecolor{YELLOW}{cmyk}{0,0,1,0}
}

\usepackage{epstopdf}
\usepackage{color}
\usepackage{array}
\usepackage{mathrsfs}
\usepackage{dcolumn}
\usepackage{bm}

\newcolumntype{L}[1]{>{\raggedright\let\newline\\\arraybackslash\hspace{0pt}}m{#1}}
\newcolumntype{C}[1]{>{\centering\let\newline\\\arraybackslash\hspace{0pt}}m{#1}}
\newcolumntype{R}[1]{>{\raggedleft\let\newline\\\arraybackslash\hspace{0pt}}m{#1}}

\graphicspath{{../Inkscape/},{../Mathematica/},{../Mathematica/Paper_figures/}}

\newcommand{\mbf}[1]{\mathbf{#1}}

\usepackage{tikz}
\usetikzlibrary{calc,intersections}
\usetikzlibrary{shadings}
\usepgflibrary{shadings}

\makeatother

\begin{document}
\title{Orbital transmutation and the electronic spectrum of FeSe in the nematic phase}
\author{Morten H. Christensen}
\email{mchrist@umn.edu}
\author{Rafael M. Fernandes}
\author{Andrey V. Chubukov}
\affiliation{School of Physics and Astronomy, University of Minnesota, Minneapolis,
Minnesota 55455, USA}
\date{\today}
\begin{abstract}
We consider the electronic spectrum near $M=(\pi,\pi)$
in the nematic phase of FeSe ($T<T_{{\rm nem}}$) and make a detailed
comparison with recent ARPES and STM experiments. Our main focus is
the unexpected temperature dependence of the excitations at the
$M$ point. These have been identified as having $xz$ and $yz$ orbital
character well below $T_{{\rm nem}}$, but remain split at $T>T_{{\rm nem}}$, in apparent contradiction to the fact that in the tetragonal phase the $xz$ and $yz$ orbitals are degenerate. Here we present two scenarios which can describe the data. In both scenarios, hybridization terms present in the tetragonal phase leads to an orbital transmutation, a change in the dominant orbital character of some of the bands, between $T > T_{\rm nem}$ and $T \ll T_{\rm nem}$. The first scenario relies on the spin-orbit coupling at the $M$ point. 
We show that a finite spin-orbit coupling gives rise to orbital transmutation, in which one of the modes, identified as $xz$ ($yz)$ at $T \ll T_{{\rm nem}}$, becomes predominantly $xy$ at $T >  T_{{\rm nem}}$ and hence does not merge with the predominantly $yz$ ($xz$) mode. The second scenario, complementary to the first, takes into consideration the fact that both ARPES and STM are surface probes. In the bulk, a direct hybridization between the $xz$ and $yz$ orbitals is not allowed at the $M$ point, however, it is permitted on the surface.
In the presence of a direct $xz/yz$ hybridization, the orbital character of the $xz/yz$ modes changes from pure $xz$ and pure $yz$ at $T \ll T_{{\rm nem}}$ to $xz \pm yz$ at $T > T_{{\rm nem}}$, i.e., the two modes again have mono-orbital character at low $T$, but do not merge at $T_{{\rm nem}}$. We discuss how these scenarios can be distinguished in polarized ARPES experiments.
\end{abstract}
\maketitle

\section{Introduction}

The intriguing physical properties of FeSe continue to attract the
attention of the correlated electron systems community~\cite{bohmer17,sprau17,coldea18}. This material
has the simplest structure among the Fe-based superconductors (FeSCs),
yet its phase diagram is rather complex, particularly under pressure~\cite{sun2016,bohmer2018},
and is quite different from that of other FeSCs. The most notable
distinction is a wide temperature range where the tetragonal symmetry
of the lattice is spontaneously broken down to $C_{2}$ (the nematic
phase). The nematic order emerges at $T_{{\rm nem}}\sim90$~K at
nominal pressure, and is not accompanied by a stripe magnetic order~\cite{mcqueen09}.
Superconductivity emerges at a much smaller $T\sim8$~K~\cite{hsu08}.

The electronic structure of FeSe in the tetragonal phase ($T>T_{{\rm nem}}$)
is fairly typical of the FeSCs -- there are two cylindrical hole pockets
centered at $\Gamma=(0,0)$ in the Brillouin zone (BZ) and two cylindrical
electron pockets~\cite{coldea18}. In the 1-Fe BZ, one of the electron pockets
is centered at $(\pi,0)$ (the $X$ point), while the other is centered
at $(0,\pi)$ (the $Y$ point)~\cite{footnote_notation}. The hole pockets consist equally of fermions from the $xz$ and
$yz$ orbitals, the $X$ pocket is made of fermions from the $yz$
and $xy$ orbitals, and the $Y$ pocket of fermions from the
$xz$ and $xy$ orbitals~\cite{coldea18}. In the crystallographic 2-Fe unit
cell, the $X$ and $Y$ pockets are
folded onto $M=(\pi,\pi)$, or, equivalently, $M'=(\pi,-\pi)$, and form inner and outer pockets~\cite{lin11}.
The inner pocket is made mostly
of $xz$ and $yz$ fermions, and the outer pocket is made
mostly of $xy$ fermions~\cite{coldea18}. The
two pockets touch along high-symmetry axes in the absence of spin-orbit coupling
(SOC) and split in its presence~\cite{cvetkovic13,christensen15,borisenko16}.  Although the same geometry of low-energy
excitations holds in other FeSCs, the peculiarity of FeSe is that
the Fermi energies for the hole and electron bands are smaller than
in other FeSCs~\cite{terashima14,audouard15,maletz14,rhodes17,kushnirenko17,watson15}.

\begin{figure}
\includegraphics[width=0.95\columnwidth]{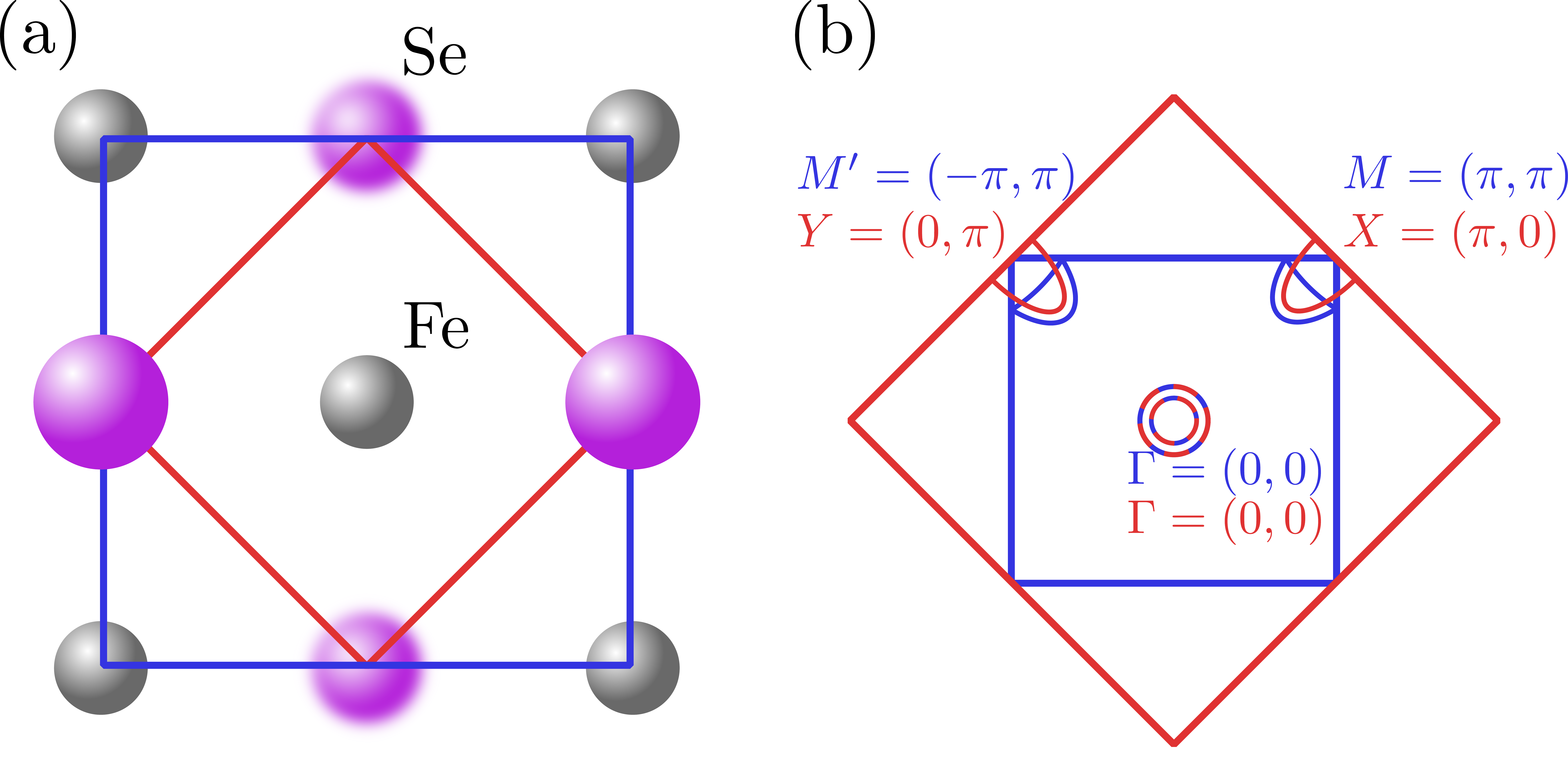}
\caption{\label{fig:unit_cell} 1-Fe and 2-Fe unit cells in (a) real and (b) momentum space. In (a) gray denotes Fe-atoms and purple denotes the Se-atoms puckered above and below the Fe-plane. The rotated red square denotes the 1-Fe unit cell, while the blue square denotes the 2-Fe unit cell. In (b) the corresponding unit-cells in momentum space are shown with schematic Fermi surfaces overlaid. In the 1-Fe unit cell (red), one electron pocket is centered at both $X$ and $Y$. In the 2-Fe unit cell (blue), there are two electron pockets at $M$ and two at $M'$ (note that the use of $M'=(-\pi,\pi)$ is not standard). In both unit cells, two hole pockets are centered at $\Gamma$.}
\end{figure}

In the nematic phase, the occupations of the $xz$ and $yz$
orbitals become inequivalent on both hole and electron pockets~\cite{shimojima14}, and the occupations of the $xy$ orbitals at $X$ and at $Y$ also generally become different (the latter gives rise to a hopping anisotropy in real space~\cite{christensen19}).
This changes both the shape of the pockets and the orbital composition
of excitations along them.  Of the two hole pockets, the smaller
one sinks below the Fermi level, and the larger one becomes elliptical~\cite{shimojima14,suzuki15,fanfarillo16}.
The direction of its longer axis (towards $X$ or $Y$ in
the 1-Fe BZ, or, equivalently, towards $M$ or $M'$ in the 2-Fe BZ) is chosen spontaneously. In non-strained (twinned)
samples, Angle Resolved Photo-Emission Spectroscopy (ARPES) measurements
necessarily see a superposition of the two hole pockets elongated
along orthogonal directions, due to the presence of twin domains~\cite{watson15}.
However, polarized ARPES measurements on twinned samples or unpolarized
measurements in detwinned samples allow one to focus on a single domain.
Here, we follow ARPES data
and focus on the domain in which, deep in the nematic phase, the hole
pocket is elongated towards $Y$ in the 1-Fe BZ~\cite{suzuki15,watson17}, or, equivalently,
 along $\Gamma-M'$ in the 2-Fe BZ  (we use the convention that $\Gamma-Y$ direction in the 1-Fe BZ corresponds to $\Gamma-M'$ in the 2-Fe BZ, see Fig.~\ref{fig:unit_cell} for the an explanation of the relationship between the 1- and 2-Fe unit cells~\cite{lin11}.)
For the electron pockets, both ARPES~\cite{suzuki15,watson16,yi19,huh19} and Scanning Tunneling Microscopy (STM) measurements~\cite{sprau17} have shown that, within the same domain, the inner electron pocket acquires a peanut-like shape, with smaller axis towards $M'$ (larger axis towards $M$).

\begin{figure}
\centering \includegraphics[width=0.95\columnwidth]{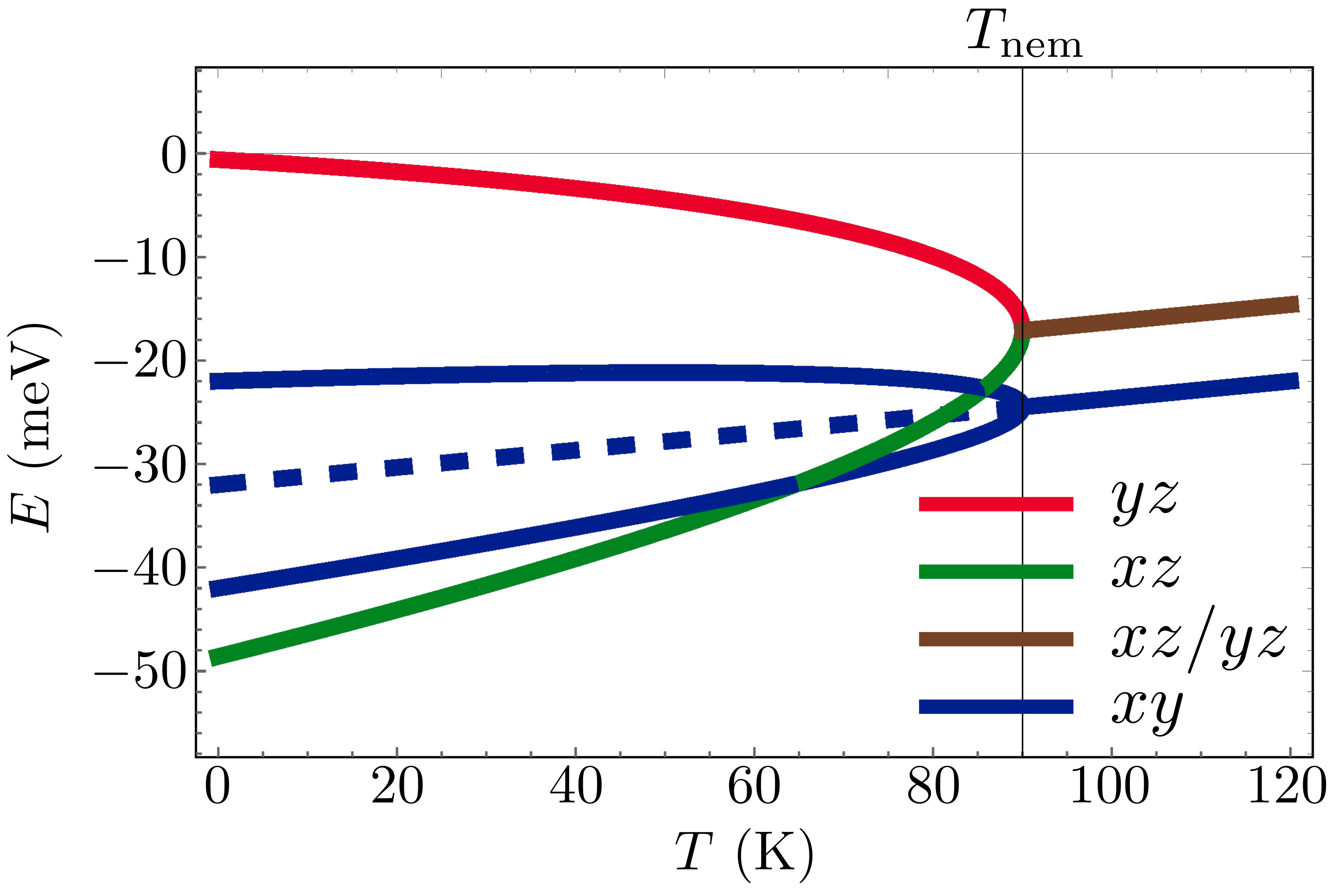}
\caption{\label{fig:standard_E} Evolution of the states at the $M$ point
as a function of temperature for the ``standard model''  of Eq.~\eqref{eq:nem_M}. The full blue lines correspond to the case with $\phi_3 = \pm 10$ meV, while the dashed blue line corresponds to the case $\phi_3 = 0$. At $T\ll T_{\mathrm{nem}}$, the four excitations are the orbital eigenstates $xz$ (green),
$xy$ (blue), and $yz$ (red). Upon approaching $T_{\mathrm{nem}}$, the energies of the $xz$ and $yz$ states merge, and the two form a doublet, as does the energies of the two $xy$ states.}
\end{figure}

\begin{figure}
\centering \includegraphics[width=\columnwidth]{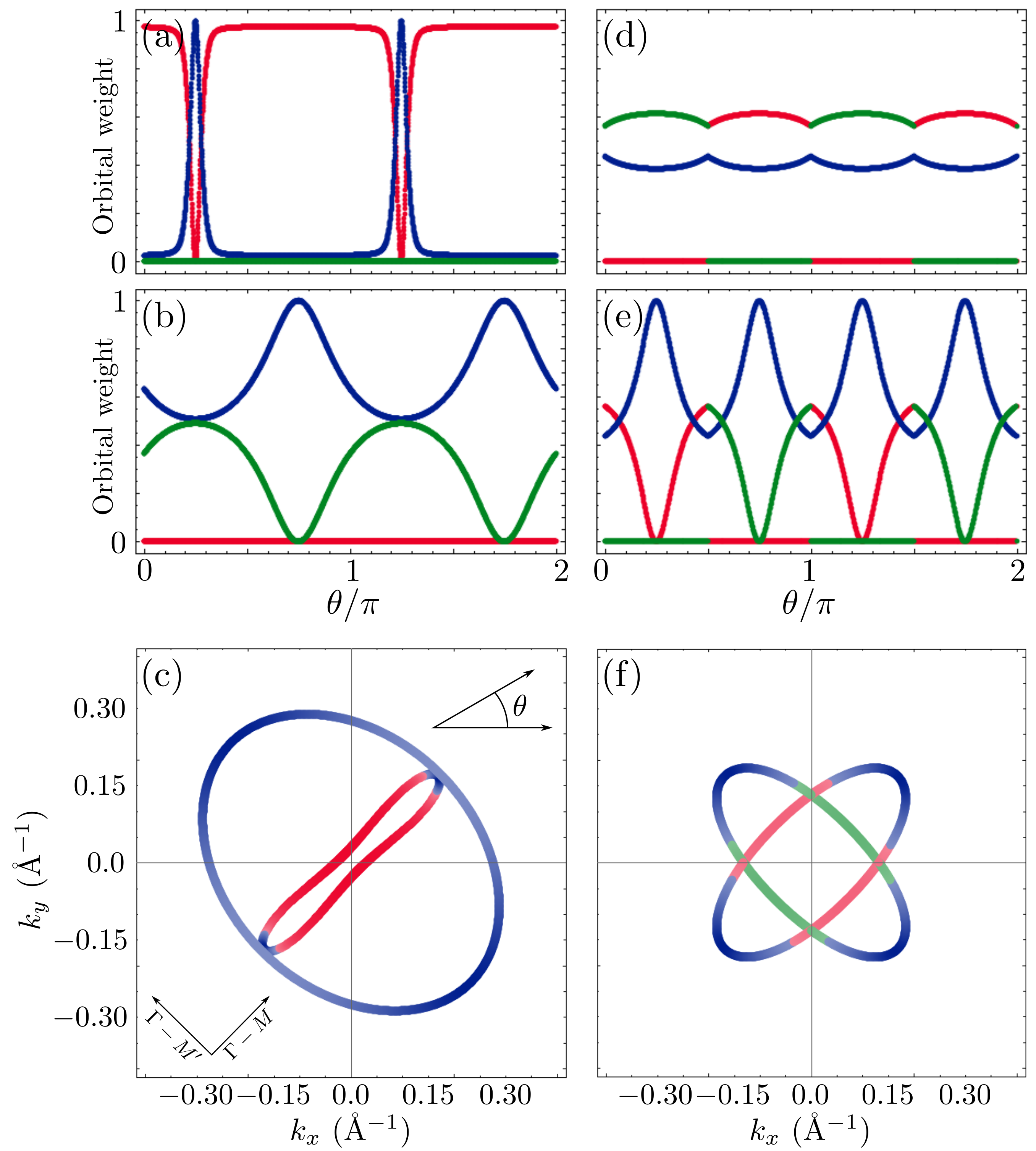}
\caption{\label{fig:standard_fs} Fermi surfaces and orbital weights at $T\ll T_{\mathrm{nem}}$
(a)--(c) and $T>T_{\mathrm{nem}}$ (d)--(f)
for the standard case in which only the terms of Eq.~(\ref{eq:nem_M}) and the $xy$ hopping anisotropy
are included ($\phi_3 =-10$ meV here). The orbital color code is the same as in Fig.~\ref{fig:standard_E}.}
\end{figure}

The change of the shapes of the hole and electron pockets can be well understood
at the mean-field level, by adding to the kinetic energy the fermionic
bilinears that couple directly to the nematic order parameters:
\begin{align}
\mathcal{H}_{{\rm nem}}^{\Gamma} & = \phi_{\Gamma}\left(\tilde{d}_{xz,\sigma}^{\dagger}\tilde{d}_{xz,\sigma}-
\tilde{d}_{yz,\sigma}^{\dagger}\tilde{d}_{yz,\sigma}\right) \nonumber \\
\mathcal{H}_{{\rm nem}}^{M} & = \phi_1(\hat{d}_{xz,\sigma}^{\dagger}\hat{d}_{xz,\sigma}-d_{yz,\sigma}^{\dagger}d_{yz,\sigma})  \label{eq:nem_M} \\
 & + \phi_3(\hat{d}_{xy,\sigma}^{\dagger}\hat{d}_{xy,\sigma}-d_{xy,\sigma}^{\dagger}d_{xy,\sigma})\,. \nonumber
\end{align}
Here $\phi_{\Gamma}$ and $\phi_{1}$ are the nematic orders
associated with the $xz$ and $yz$ orbitals near $\text{\ensuremath{\Gamma}}$
and $M$, respectively, and $\phi_{3}$ is the nematic order
associated with the two $xy$ orbitals near $M$~\cite{vafek14}. We use $\tilde {d}$ for the states at $\Gamma$, $\hat{d}$ for the states at the
$Y$ point in the 1-Fe BZ and ${d}$ for the states at the $X$ point.
The elongation of the hole pocket along $\Gamma-{M}'$ and the
peanut-like form of the electron pocket, with larger axis along
$\Gamma-{M}$, are reproduced if $\phi_{\Gamma}>0$
and $\phi_1<0$ (Refs. \cite{sprau17,kontani_1,kang18,latest}). We will use this convention throughout. It implies that the $xz$ orbital is the dominant one for the hole pocket, while the $yz$ orbital is the dominant one for the inner electron pocket. The sign change between $\phi_{\Gamma}$ and $\phi_1$
is consistent with the theoretical reasoning that, for repulsive interactions, a spontaneous nematic order is possible only if it changes sign between hole and electron pockets~\cite{chubukov16}, similarly to an $s^{+-}$ superconducting order.
The $\phi_3$ term leads to a splitting of the two degenerate $xy$ orbitals from the $X$ and $Y$ pockets (see Fig.~\ref{fig:standard_E}). This term does not affect the peanut-like shape (shown in Fig.~\ref{fig:standard_fs}) of the inner electron pocket~\cite{christensen19} and will play only a secondary role in our analysis. The sign (and magnitude) of this term has not been verified in experiments. In renormalization group calculations~\cite{xing16} the sign is the same as of $\phi_1$.

While the shape of the pockets near $M$ can be well understood by including only the standard terms in Eq.~\eqref{eq:nem_M}, the ARPES data for the excitations right at the $M$ point cannot be straightforwardly
explained within the ``standard model'' of Eq.~\eqref{eq:nem_M}.
Specifically, above $T_{\mathrm{nem}}$, the $yz$ and $xz$
states at $M$ form a doublet and are degenerate. Below $T_{\mathrm{nem}}$
they split exactly by $2\phi_{1}$, according to Eq.~\eqref{eq:nem_M}
(see Fig.~\ref{fig:standard_E}). ARPES measurements deep in the nematic
state do detect two sharp excitations~\cite{watson15,watson16,fedorov16,watson17,rhodes17,rhodes18,huh19,yi19}, and a recent polarized ARPES study~\cite{yi19} provided strong evidence that these two states are indeed $xz$- and $yz$-dominated, by tracking them from the $M$ to the $\Gamma$ point.
However, as $T$ increases towards $T_{{\rm nem}}$, the two states
do not merge and remain split even above $T_{{\mathrm{nem}}}$~\cite{watson15,watson16,fedorov16,watson17,rhodes17,rhodes18,huh19,yi19}. In Ref.~\onlinecite{yi19} it was speculated that the absence of merging may be a spurious feature due to thermal broadening, while in Refs.~\onlinecite{watson15,watson16,fedorov16,watson17,rhodes17,rhodes18} it is argued that the splitting of the two modes above $T_{\rm nem}$ is a physical feature. If this is indeed the case, the ARPES data at $M$ are inconsistent with the standard model of Eq.~\eqref{eq:nem_M}.

In this communication we argue that the ARPES data at the $M$ point
can be explained if one includes SOC for the fermions near the electron pockets. In the presence of SOC, the excitations at $M$ are no longer orbital eigenstates. In the tetragonal phase this does not lead to  drastic changes: the excitations still form two doubly degenerate states, such that the states in the doublet closer to the Fermi level are $xz$ and $yz$ with small admixtures of $xy$, while the states in the doublet farther from the Fermi level are $xy$, one with a small admixture of $xz$, the other with a small admixture of $yz$ (see Fig.~\ref{fig:evolution_at_M_with_T_soc}). Our key finding is that below $T_{\mathrm{nem}}$, where the double degeneracy is lost, an excitation from each doublet undergoes an \emph{orbital transmutation},
i.e. its dominant orbital contribution changes compared to that in the tetragonal phase. Specifically, the doublet closer to the Fermi level splits into two energy levels: the upper one remains predominantly $yz$, while the lower one becomes predominantly of $xy$ character  at $T\ll T_{\mathrm{nem}}$. The other doublet also splits into two levels. In this case, the upper one remains predominantly $xy$, while the lower one becomes predominantly of $xz$ character at $T\ll T_{\mathrm{nem}}$. As $T$ increases towards $T_{\mathrm{nem}}$, the $yz$ excitation remains
sharp and becomes a part of the upper doublet in the tetragonal
phase. The excitation dominated by the $xz$ orbital at $T \ll T_{\rm nem}$ becomes more incoherent as temperature is increased and the dominant orbital weight changes from $xz$ to $xy$. At $T_{\rm nem}$ this excitation merges with the lower, $xy$ dominated doublet.  We show the excitations in Fig.~\ref{fig:evolution_at_M_with_T_soc} (for different values of $\phi_{3}$) and the spectral function in Fig.~\ref{fig:soc_spectral_function_at_M}.

The orbital transmutation
can be gleaned by looking at the excitations in the standard model of Eq.~\eqref{eq:nem_M}. From Fig.~\ref{fig:standard_E} we see that the $xz$ band crosses the $xy$ excitations. In the absence of SOC, this is just a level crossing
 as $xz$ and $xy$ orbitals are not allowed to hybridize at the $M$ point.  A non-zero SOC gives rise to level repulsion between the $xz$ excitation and one of the $xy$ excitations (the one from the $X$ pocket in the 1-Fe BZ), see Eq.~\eqref{eq:soc}. As a result, the orbital weight is transferred between the two excitations.
We show that a similar behavior emerges if instead of SOC we include a hybridization between the $xz$ and $yz$ orbitals at the $M$ point.
Such hybridization  is forbidden in the bulk by glide-plane symmetry~\cite{cvetkovic13}, but is allowed on the surface~\cite{christensen19} and in this regard should be viewed as surface-induced
hybridization (SIH)~\cite{footnote_sih}.  It is relevant to ARPES and STM experiments as both probe electrons near the surface. Due to SIH, the $xz/yz$ doublet is split already in the tetragonal phase into higher and lower energy excitations, with equal mixtures of $xz$ and $yz$ orbital characters (Fig.~\ref{fig:evolution_at_M_with_T_a2u}). In the nematic
phase, the orbital character of the excitation closer to the Fermi level becomes predominantly $yz$, while the orbital character of the excitation farther from the Fermi level becomes predominantly $xz$. In this
case, the modes are sharp both at  $T\ll T_{\mathrm{nem}}$ and $T> T_{\mathrm{nem}}$, but they do not merge at $T>T_{\mathrm{nem}}$ (Fig.~\ref{fig:evolution_at_M_with_T_soc_a2u}).  When both SOC and SIH are present, the two excitations should remain visible to ARPES at all temperatures, see Fig.~\ref{fig:soc_a2u_spectral_func_at_M}.

We believe that this theoretical scenario solves the puzzle of ARPES data at the $M$ point in FeSe. As we said before, all ARPES experiments observe two sharp excitations at $M$ at $T\ll T_{\mathrm{nem}}$~\cite{watson15,watson16,fedorov16,watson17,rhodes17,rhodes18,huh19,yi19}. Our results agree with Refs.~\onlinecite{watson15,watson16,fedorov16,rhodes17,rhodes18,huh19}, which followed these two excitations from $T\ll T_{\mathrm{nem}}$ to $T> T_{\mathrm{nem}}$ and argued that they remain split at $T>T_{\mathrm{nem}}$.
Our results also agree with Ref.~\onlinecite{fedorov16}, which identified two additional, less coherent excitations at $T\ll T_{\mathrm{nem}}$, located in between the
$xz$ and $yz$ dominated excitations. In our scenario, these excitations are identified as having predominantly $xy$ orbital character.

We also discuss the orbital composition of the pockets. The polarized ARPES and STM measurements show that the nematic order drastically changes the orbital content of the pockets deep in the nematic phase. In one domain the elliptical hole pocket becomes predominantly $xz$, and the inner electron pocket becomes predominantly $yz$. This is in sharp contrast to the behavior in the tetragonal phase, where the orbital content of these pockets oscillates between $xz$ and $yz$ (see Fig.~\ref{fig:standard_fs}). Such a drastic change
of the orbital content is not expected in a generic FeSC, where $\phi_{\Gamma}$
and $\phi_1$ are much smaller than the corresponding Fermi energies
$E_{F}$, but has been reproduced theoretically for FeSe~\cite{kreisel17,kang18}, where the
pockets are smaller than in other FeSCs, and $\phi_{\Gamma}$ and
$\phi_1$ are comparable to $E_{F}$. For the same parameters, calculations show that the outer
electron pocket becomes larger and more circular in the nematic phase~\cite{fanfarillo16,rhodes17,kreisel17,kang18,benfatto18},
and its orbital content is predominantly $xy$ (see Fig.~\ref{fig:standard_fs}).
In other words, all pockets become nearly mono-orbital deep in the nematic state:
the hole pocket becomes $xz$ ($yz$), the inner electron
pocket becomes $yz$ ($xz$), and the outer electron pocket
becomes $xy$ (and, if the second, smaller hole pocket does not sink completely below the Fermi level, its orbital content becomes  $yz$ ($xz$)).

The near-$xy$ composition of the outer electron pocket may explain why this pocket has not been detected
in ARPES measurements. Namely, it has been argued in several papers~\cite{lanata13,medici14,sprau17,kreisel17,yu17}, that the $xy$ fermions are either completely incoherent, or have
only a small coherent spectral weight at low-energies ($Z_{xy}\ll1$), with the rest of the
spectral weight transferred to higher energies. If this is the case,
then the pocket made of $xy$ fermions is almost invisible to
ARPES, as the large incoherent background would mask any shallow peaks.

The rest of the paper is organized as follows. In Sec.~\ref{sec:LEM}
we introduce the low-energy model and describe in more detail the
standard coupling to the nematic order parameter, the SOC for fermions
near $M$, and the SIH term. We set the parameters in the fermionic
dispersion to match the observed forms of hole and electron pockets
in the tetragonal and the nematic phases. In Sec.~\ref{sec:soc} we discuss
 the orbital composition of the Fermi pockets and
the
excitations at $M$ in the presence of SOC. In Sec.~\ref{sec:a2u}
we discuss the same in the presence of SIH. We discuss the
results and present our conclusions in Sec. \ref{sec:conclusions}.

\section{The low-energy model}

\label{sec:LEM}

We work in the crystallographic 2-Fe BZ, in which
the electron pockets at $X$ and $Y$ are folded onto the $M$ (and $M'$) points,
and form inner and outer electron pockets.
To obtain the fermionic dispersion, we follow Ref.~\onlinecite{cvetkovic13} and use a $\mbf{k}\cdot\mbf{p}$-expansion around the $M$ point, which respects all the symmetries of a single FeSe layer. In the absence of SOC, the orbital states at the $M$ point are pure eigenstates. Since the space group of FeSe, $P4/nmm$, is non-symmorphic, all irreducible representations at the $M$ point are two-fold degenerate~\cite{cvetkovic13}, in which case the excitations form doublets (quadruplets, if we include spin degeneracy). The doublet
closest to the Fermi level consists entirely of the $xz$ and $yz$ orbitals, whereas the lower doublet is made of the two $xy$ orbitals originating from the two inequivalent Fe sites in the 2-Fe BZ (the states near $X$ and near $Y$ in the 1-Fe BZ). A suitable basis for these doublets is:
\begin{eqnarray}
\Psi(\mbf{k})=\begin{pmatrix}\Psi_{Y}(\mbf{k})\\
\Psi_{X}(\mbf{k})
\end{pmatrix}\,,
\end{eqnarray}
where
\begin{eqnarray}
\Psi_{Y}(\mbf{k})=\begin{pmatrix}\hat{d}_{xz,\sigma}(\mbf{k})\\
\hat{d}_{xy,\sigma}(\mbf{k})
\end{pmatrix}\,,\quad\Psi_{X}(\mbf{k})=\begin{pmatrix}d_{yz,\sigma}(\mbf{k})\\
d_{xy,\sigma}(\mbf{k})
\end{pmatrix}\,,
\end{eqnarray}
and ${\bf k}$ is the deviation from $M$. As in Eq.~\eqref{eq:nem_M}, we label the states which are associated with the $Y$ pocket in the 1-Fe BZ point by $\hat{d}$. The kinetic energy term in this basis is:
\begin{eqnarray}
\mathcal{H}_{\rm kin} (\mbf{k})=\begin{pmatrix}h_{+}(\mbf{k}) & 0\\
0 & h_{-}(\mbf{k})
\end{pmatrix}\,,\label{eq_H}
\end{eqnarray}
with
\begin{eqnarray}
 &  & h_{\pm}(\mbf{k})=\nonumber \\
 &  & \begin{pmatrix}\epsilon_{1}+\frac{\mbf{k}^{2}}{2m_{1}}\pm a_{1}k_{x}k_{y} & -iv_{\pm}(\mbf{k})\\
iv_{\pm}(\mbf{k}) & \epsilon_{3}+\frac{\mbf{k}^{2}}{2m_{3}}\pm a_{3}k_{x}k_{y}
\end{pmatrix}\otimes\sigma^{0}\,,
\end{eqnarray}
where $\sigma^0$ is $2 \times 2$ identity matrix (we use $\boldsymbol{\sigma}$ to denote matrices in spin-space) and $v_{\pm}(\mbf{k})$ is a polynomial odd in powers of $k$:
\begin{eqnarray}
v_{\pm}(\mbf{k}) & = & v(\pm k_{x}+k_{y})+p_{1}(\pm k_{x}^{3}+k_{y}^{3})\nonumber \\
 & + & p_{2}k_{x}k_{y}(k_{x}\pm k_{y})\,.
\end{eqnarray}
At the $M$ point, ${\mbf k}=0$, the Hamiltonian is diagonal
\begin{eqnarray}
h_{+}(0)=h_{-}(0)=
\begin{pmatrix}\epsilon_{1} & 0\\
0 & \epsilon_{3}
\end{pmatrix}\otimes\sigma^{0}\,,
\end{eqnarray}
i.e., the eigenstates are pure orbital states. We assume that the onsite energies $\epsilon_{1}$ and $\epsilon_{3}$ are slowly-varying functions of temperature and use $\epsilon_{1}(T)=\epsilon_{1,0}+0.083T,~\epsilon_{3}(T)=\epsilon_{3,0}+0.083T$, where $\epsilon_{1,0}=-24.6$ meV and $\epsilon_{3,0}=-32.0$ meV, to reproduce the ARPES data at $M$ in the
tetragonal phase~\cite{suzuki15,watson15,fanfarillo16,coldea18,rhodes18,kushnirenko18}. We use these values in  Figs.~\ref{fig:standard_E} and \ref{fig:standard_fs} to obtain the variation of the excitations at $M$ in the nematic phase within the standard model. We will adjust $\epsilon_{1,0}$ and $\epsilon_{3,0}$ slightly in the presence of SOC and SIH to maintain the peanut-like shape of the inner electron pocket. The exact values will be given in the appropriate sections, but the variation between cases is rather small, within $3$ meV.

Away from the $M$ point, the eigenstates are no longer pure orbital states.
We use the parameters listed in Table~\ref{tab:parameters} to reproduce the ARPES and STM data for the peanut-shape inner electron pocket
~\cite{suzuki15,watson15,fanfarillo16,sprau17,
coldea18,rhodes18,kushnirenko18}.
\begin{table*}
\begin{tabular}{|c|c|c|c|c|c|c|c|c|}
\hline
$\epsilon_{1,0}$ & $\epsilon_{3,0}$ & $v$ & $p_1$ & $p_2$ & $(2m_1)^{-1}$ & $(2m_3)^{-1}$ & $a_1$ & $a_3$ \\
\hline
$-24.6$ & $-32.0$ & $-122.90$ & $-137.22$ & $-11.67$ & $1.41$ & $186.11$ & $136.12$ & $-403.84$ \\
\hline
\end{tabular}
\caption{\label{tab:parameters} Parameters used in the low-energy model listed in units of meV.}
\end{table*}
The standard coupling of fermions near the $M$ point to the nematic order parameters $\phi_1$ and $\phi_3$ is given by Eq.~\eqref{eq:nem_M}. For our analysis it suffices to consider a mean-field temperature-dependence
of $\phi_1$ and $\phi_3$: $\phi_{i}(T)= \phi_{i,0}\sqrt{1-T/T_{\mathrm{nem}}}$, where $T_{\mathrm{nem}}=90$ K. For definiteness, we set $\phi_{1,0}=-24$ meV in all calculations, and use three different values of $\phi_{3,0}$ ($\phi_{3,0} = 0$ and $\phi_{3,0} = \pm 10$ meV). We will see that the results do not depend substantially on the choice of $\phi_{3,0}$.

The Fermi surfaces and the thermal evolution of the excitations at the $M$-point within this standard model are shown in Figs. \ref{fig:standard_E} and \ref{fig:standard_fs}. The shape of the inner pocket and its orbital composition are consistent with ARPES and STM data. The outer electron pocket has predominantly $xy$ orbital character. Fermions on the $xy$ orbital are likely incoherent~\cite{lanata13,medici14}, which may explain why this pocket has not been observed in ARPES studies in the nematic phase.

We see from Fig.~\ref{fig:standard_E} that the excitations at $M$ retain their orbital character in the nematic phase: two of the four modes are $xy$, one is $xz$, and one is $yz$.
The  $xz$ and $yz$  excitations are split at $T<T_{\mathrm{nem}}$, but merge at $T =T_{\mathrm{nem}}$ and form a doublet at $T>T_{\mathrm{nem}}$
(the energies of the two $xy$ modes merge into another doublet).
As stated above, this is inconsistent with ARPES. It is argued in several papers~\cite{watson15,watson16,watson17,rhodes17,rhodes18,fedorov16,huh19}, that the excitations identified as $xz$ and $yz$ deep in the nematic state remain split at $T=T_{\mathrm{nem}}$. In the next Section we show that a
new effect, dubbed orbital transmutation, emerges once we extend the
standard model to include SOC, and the evolution of the excitations
in the model with SOC is fully consistent with ARPES data.

The form of the SOC at the $M$ point has been analyzed in Ref.~\onlinecite{cvetkovic13} using symmetry considerations. It was argued that SOC leads to an additional off-diagonal term in the Hamiltonian $\mathcal{H}_{\rm kin}(\mbf{k})$ in Eq.~\eqref{eq_H}:
\begin{eqnarray}
\mathcal{H}_{{\rm SOC}} =\begin{pmatrix}
0 & h_{{\rm SOC}}\\
h_{{\rm SOC}}^{\dagger} & 0
\end{pmatrix}\,,\label{eq_H_1}
\end{eqnarray}
where
\begin{eqnarray}
h_{{\rm SOC}} & = & \frac{i}{2}\lambda\left[\begin{pmatrix}0 & 1\\
0 & 0
\end{pmatrix}\otimes\sigma^{1}+\begin{pmatrix}0 & 0\\
1 & 0
\end{pmatrix}\otimes\sigma^{2}\right]\,.
\end{eqnarray}
In terms of the operators at $M$ this is
\begin{eqnarray}
\mathcal{H}_{{\rm SOC}} & = & \frac{i}{2}\lambda\,\hat{d}_{xz,\alpha}^{\dagger}\sigma_{\alpha\beta}^{x}d_{xy,\beta}\nonumber \\
 & + & \frac{i}{2}\lambda\,\hat{d}_{xy,\alpha}^{\dagger}\sigma_{\alpha\beta}^{y}d_{yz,\beta}+\text{H.c.}\,.\label{eq:soc}
\end{eqnarray}
In the 1-Fe BZ, this couples $xz$ fermions at $Y$ with
$xy$ fermions at $X$, and $yz$ fermions at $X$
with $xy$ fermions at $Y$. We emphasize that the coupling in Eq. (\ref{eq:soc}) is the only symmetry allowed momentum-independent SOC. Momentum-independent couplings between $\hat{d}_{xz,\sigma}$ and $\hat{d}_{xy,\sigma}$ or $d_{yz,\sigma}$ and $d_{xy,\sigma}$, i.e. between fermions from the same electron pocket in the 1-Fe BZ, are not allowed by symmetry~\cite{cvetkovic13}. Momentum-dependent SOC terms are allowed~\cite{cvetkovic13,eugenio18}, but these terms are expected to be small in FeSe because the relevant momenta are small.

\begin{figure}
\includegraphics[width=0.95\columnwidth]{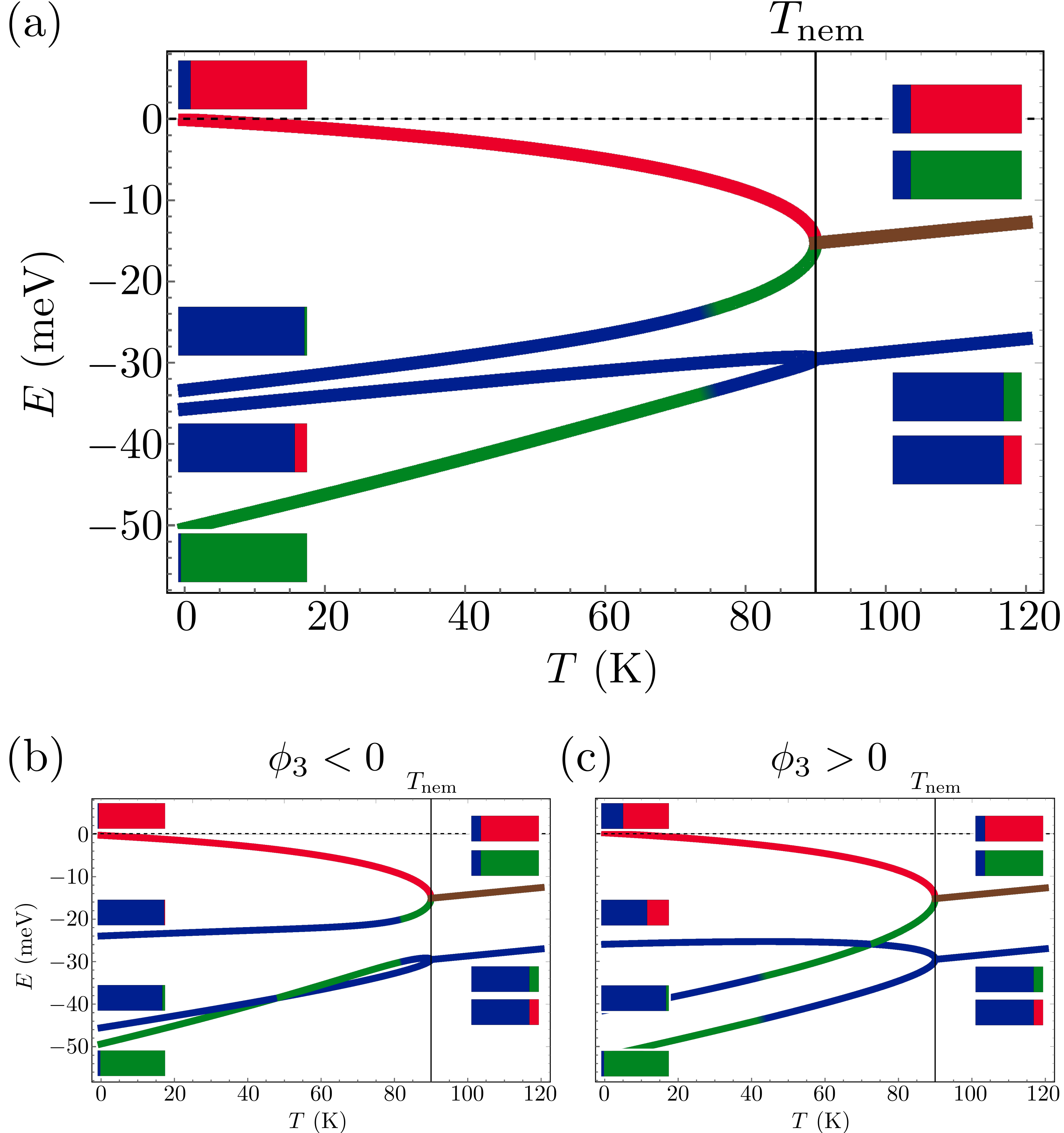}
\caption{\label{fig:evolution_at_M_with_T_soc} Evolution of the energies of the $M$-point excitations as a function of temperature for the SOC
scenario with $\lambda = 10$ meV. The orbital color code is the same as Fig.~\ref{fig:standard_E}. Here $\epsilon_{1,0}=-24.8$ and $\epsilon_{3,0} = -35.0$. In (a) we show the situation where $\phi_3 =0$ while in (b) and (c) we illustrate the difference between $\phi_3=-10$ meV (b) and $\phi_3 =10$ meV (c). The inclusion of SOC hybridizes the $xz/yz$ states with $xy$ states leading to an orbital transmutation at lower temperatures, such that the state with predominantly $xz$ character at $T\ll T_{{\rm nem}}$ originates from the lower of the two doublets at the $M$ point. The colored bars denote the orbital weight at respectively 0 K and 120 K for each of the excitations.}
\end{figure}

The sum of the kinetic energy term and the SOC term is
\begin{eqnarray}
\mathcal{H}_{\rm tot}(\mbf{k})=\begin{pmatrix}h_{+}(\mbf{k}) & h_{{\rm SOC}}\\
h_{{\rm SOC}}^{\dagger} & h_{-}(\mbf{k})
\end{pmatrix}\,,\label{eq_H_2}
\end{eqnarray}
In the presence of SOC, excitations in the tetragonal phase still form two doublets (two quadruplets if we include spin), but each eigenstate no longer has pure orbital character (see Fig.~\ref{fig:evolution_at_M_with_T_soc} at $T > T_{\mathrm{nem}}$). The two states from the upper doublet are still dominated by the $xz$ and $yz$ orbitals, but each now has an admixture of the $xy$ orbitals. Similarly, the two states from the lower doublet are still primarily $xy$, but one has an admixture of $xz$ and the other of $yz$. In the next Section we analyze how these states evolve in the nematic phase.

The SOC in Eq.~\eqref{eq_H_1} is the bulk term, and its form is set by the symmetry of the $P4/nmm$ space group of the bulk. In Sec. IV we consider the special situation at the surface, as both ARPES and STM generally probe electrons near the surface. The glide-plane symmetry is broken at the surface, which allows additional coupling terms to be present. One can describe the effect of the surface in terms of an effective electric field
$\eta$, perpendicular to the surface \cite{Kang16,Agterberg17}. Such a field transforms as the $A_{2u}$ irreducible representation of the $P4/nmm$ space group at the zone center. Consequently, it couples to the fermionic
bilinear at the $M$-point, $\hat{d}_{xz,\sigma}^{\dagger}d_{yz,\sigma}+\text{H.c.}$, which also transforms as $A_{2u}$. As a result, the Hamiltonian for fermions at the surface acquires an additional term
\begin{eqnarray}
\mathcal{H}_{{\rm surf}}=\eta\,\hat{d}_{xz,\sigma}^{\dagger}d_{yz,\sigma}+\text{H.c.}\,.\label{SIH}
\end{eqnarray}
In the presence of such term, the $xz$ and $yz$ orbitals
remain degenerate at the $M$ point in the tetragonal phase, but the
new eigenstates are $d_{\pm,\sigma}=(\hat{d}_{xz,\sigma}\pm d_{yz,\sigma})/\sqrt{2}$.
This hybridization term splits the $xz/yz$ doublet and gives
rise to two distinct states with mixed $xz/yz$ character
at $M$ (see Fig.~\ref{fig:evolution_at_M_with_T_a2u} at $T > T_{\mathrm{nem}}$).

\section{Excitations in the nematic phase in the presence of SOC}
\label{sec:soc}

\begin{figure}
\includegraphics[width=0.95\columnwidth]{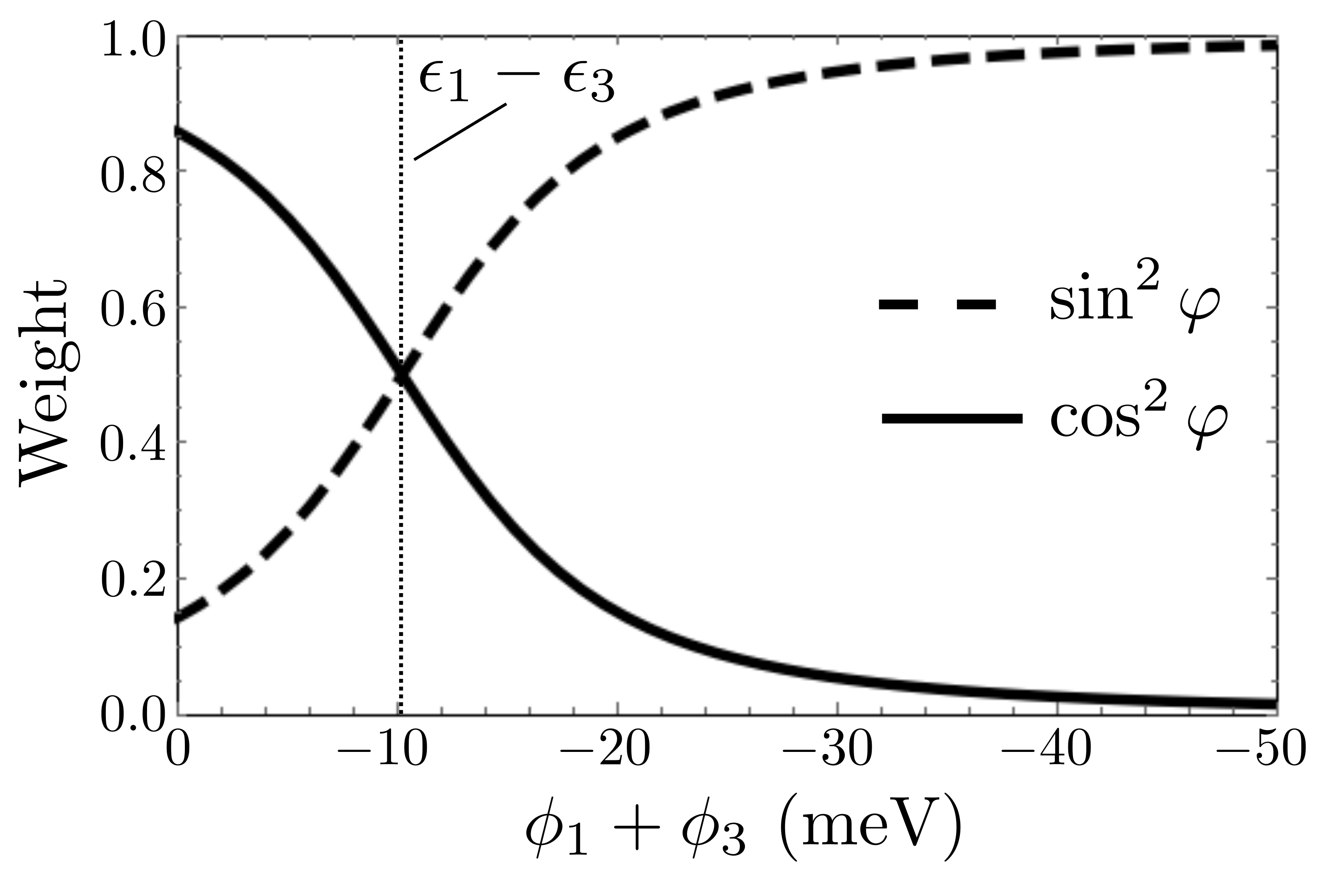}
\caption{\label{fig:orb_evolution_analytical}Evolution of $\cos^2\varphi$ and $\sin^2 \varphi$, given by ~\eqref{eq:phi_eq}, with increasing nematic order parameter $\phi_1 + \phi_3$. We used $\epsilon_1 = -25$ meV, $\epsilon_3=-35$ meV and $\lambda = 10$ meV. As discussed in the text, when the magnitude of $\phi_1 + \phi_3$ exceeds $\epsilon_1 - \epsilon_3$, the dominant orbital character of the two excitations at $M$ flips. When $\phi_1 + \phi_3$ increases further, the excitations become increasingly mono-orbital.}
\end{figure}

The full Hamiltonian at $M$ in the nematic phase is the sum of $\mathcal{H}_{{\rm nem}}^{M}$ from Eq.~\eqref{eq:nem_M}
and $\mathcal{H}_{\rm tot}(\mbf{k})$ from Eq.~\eqref{eq_H_2}. Because the atomic SOC only couples $\hat{d}_{xz}$ with $d_{xy}$ and $d_{yz}$ with $\hat{d}_{xy}$
the Hamiltonian matrix at the $M$ point decouples into four $2 \times 2$ matrices. Each matrix can be straightforwardly diagonalized.
Consider, e.g., $\hat{d}_{xz,\uparrow}$ and $d_{xy,\downarrow}$. The corresponding $2\times 2$ matrix Hamiltonian is
\begin{eqnarray}
	\begin{pmatrix}
		\epsilon_1 + \phi_1 & \frac{i}{2} \lambda \\
		-\frac{i}{2} \lambda & \epsilon_3 - \phi_3
	\end{pmatrix}\,.
\label{eq:2by2_hamiltonian}
\end{eqnarray}
The eigen-energies are
\begin{eqnarray}
	E_{\pm} &=& \frac{1}{2}\left(\epsilon_1 +\epsilon_3 +\phi_1 - \phi_3 \pm \sqrt{\lambda^2 + (\epsilon_1 - \epsilon_3 +\phi_1 + \phi_3)^2} \right)\,, \nonumber \\
&&
\end{eqnarray}
and the transformation to the diagonal basis is
\begin{eqnarray}
 && \mathfrak{a} = \hat{d}_{xz,\uparrow} \cos{\varphi} -i  d_{xy,\downarrow} \sin{\varphi} \nonumber \\
 && \mathfrak{b} = d_{xy,\downarrow} \cos{\varphi} -i \hat{d}_{xz,\uparrow} \sin{\varphi}\,,
\label{eq:diag_basis}
\end{eqnarray}
where
\begin{equation}
\tan{2\varphi} = -\frac{\lambda}{\epsilon_1 - \epsilon_3 +\phi_1 + \phi_3}\,.
\label{eq:phi_eq}
\end{equation}
Accordingly, $E_{+}$ is the energy of the $\mathfrak{a}$ excitation while $E_{-}$ is the energy of the $\mathfrak{b}$ excitation. In the absence of SOC and for $\phi_{1,3} = 0$, we find $\varphi=0$. In this case, $E_{+}=\epsilon_1$, and the excitation is made exclusively by $xz$ fermions,  while $E_{-} = \epsilon_3$ is made by $xy$ fermions. For finite SOC, the weight of the $xz$ component for the $\mathfrak{a}$ fermions
is $\cos^{2} \varphi$ and the weight of
the $xy$ component is $\sin^2 \varphi$. For the $\mathfrak{b}$ fermions
the situation is the opposite: $\cos^2\varphi$ is the weight
of the $xy$ component, while $\sin^2 \varphi$ is the weight of
$xz$. In Fig.~\ref{fig:orb_evolution_analytical} we show the evolution of $\varphi$ with increasing the magnitude of $\phi_1 + \phi_3$.

For $d_{yz,\uparrow}$ and $\hat{d}_{xy,\downarrow}$ (or $d_{yz,\downarrow}$ and $\hat{d}_{xy,\uparrow}$) the excitation energies are instead
\begin{eqnarray}
	\bar{E}_{\pm} &=& \frac{1}{2}\left(\epsilon_1 +\epsilon_3 -\phi_1 + \phi_3 \pm \sqrt{\lambda^2 + (\epsilon_1 - \epsilon_3 -\phi_1 - \phi_3)^2} \right), \nonumber \\
&&
\end{eqnarray}
and the transformation to the band basis is
\begin{eqnarray}
	&& \bar{\mathfrak{a}} = d_{yz,\uparrow} \cos{{\bar \varphi}} - {\hat d}_{xy,\downarrow} \sin{{\bar \varphi}} \nonumber \\
 && \bar{\mathfrak{b}} = {\hat d}_{xy,\downarrow} \cos{{\bar \varphi}} + d_{yz,\uparrow} \sin{{\bar \varphi}}\,,
\end{eqnarray}
with
\begin{eqnarray}
	\tan 2\bar{\varphi} = -\frac{\lambda}{\epsilon_1 - \epsilon_3 - \phi_1 - \phi_3}\,. \label{eq:phi_bar_eq}
\end{eqnarray}
In this case, the weight of the $yz$ orbital for the $\bar{\mathfrak{a}}$ excitation (the one with energy $E_{+}$) is $\cos^2 \bar{\varphi}$, while the weight of $xy$ is $\sin^2\bar{\varphi}$. For the $\bar{\mathfrak{b}}$ excitation the weight of $xy$ is $\cos^2 \bar{\varphi}$, and the weight of $yz$ is $\sin^2 \bar{\varphi}$.

The key to the phenomenon of orbital transmutation is the fact that in FeSe the difference $\epsilon_1-\epsilon_3$ is smaller than in other Fe-based materials (recall $\epsilon_3 < \epsilon_1 < 0$). At some $T$ below
$T_{\mathrm{nem}}$, the combined value of the nematic order parameters, $\phi_1 + \phi_3 = -|\phi_1| + \phi_3$ becomes larger in magnitude than $\epsilon_1-\epsilon_3$ (recall $\phi_1 < 0$ by assumption)~\cite{footnote_other_domain}. This can actually be seen from  Fig.~\ref{fig:standard_E}: the $xz$ band, whose energy is
 $\epsilon_1 + \phi_1$ in the absence of SOC, crosses both $xy$ excitations, whose energies are $\epsilon_3 \pm \phi_3$. At low $T$, $\epsilon_1- \epsilon_3 + \phi_1 + \phi_3$ therefore changes sign and becomes negative.
This implies that $\varphi$, given by Eq.~\eqref{eq:phi_eq}, varies from $\varphi \approx -0$ to $\varphi \approx - \frac{\pi}{2}$.
Accordingly, the content of the $\mathfrak{a}$ excitation at $E_{+}$ varies from near-$xz$ at $T \gtrsim T_{\rm nem}$ to near-$xy$ at $T \ll T_{\rm nem}$. Similarly, the content of the $\mathfrak{b}$ excitation varies from near-$xy$ in the high-temperature regime to near-$xz$ deep in the nematic phase (see Fig.~\ref{fig:orb_evolution_analytical}).
For the $\bar{\mathfrak{a}}$ and $\bar{\mathfrak{b}}$ excitations no orbital transmutation occurs because $\epsilon_1 - \epsilon_3 -\phi_1 - \phi_3$ in the denominator of Eq.~\eqref{eq:phi_bar_eq} does not change sign.

\begin{figure}
\centering \includegraphics[width=0.95\columnwidth]{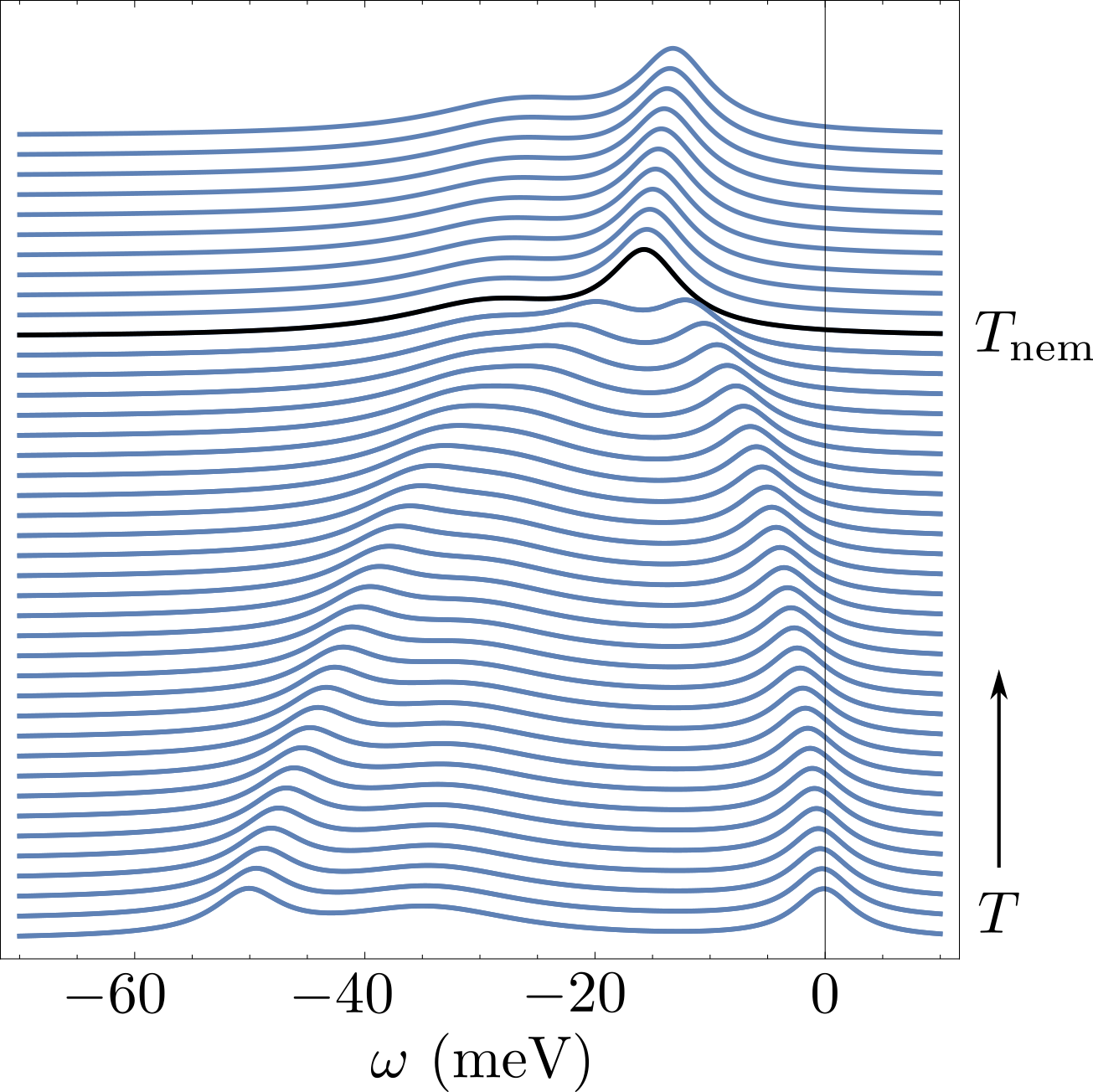} \caption{\label{fig:soc_spectral_function_at_M} Spectral function as a function
of $T$ and $\omega$ at the $M$ point.
We set damping rates to be $\Gamma_{xy}=10$~meV, and $\Gamma_{xz}=\Gamma_{yz}=3$~meV.
One peak is clearly seen
above $T_{{\rm nem}}$, originating from the $xz/yz$ doublet. Another
shallower peak from the $xy$ states can be seen farther from the
Fermi level. At $T\ll T_{{\rm nem}}$ two peaks can be easily distinguished,
the one closest to $\omega=0$ tracing back to the strong peak at
high temperatures. This is in contrast to the peak at $\omega\approx-50$
meV, which traces back to the shallower, $xy$ dominated peak at higher
temperatures. In addition, there is a broad peak at $\omega\approx-25$
meV with predominantly $xy$ character, while the fourth peak is masked
by the peak at $\omega\approx-50$ meV.}
\end{figure}

\begin{figure}
\centering \includegraphics[width=1\columnwidth]{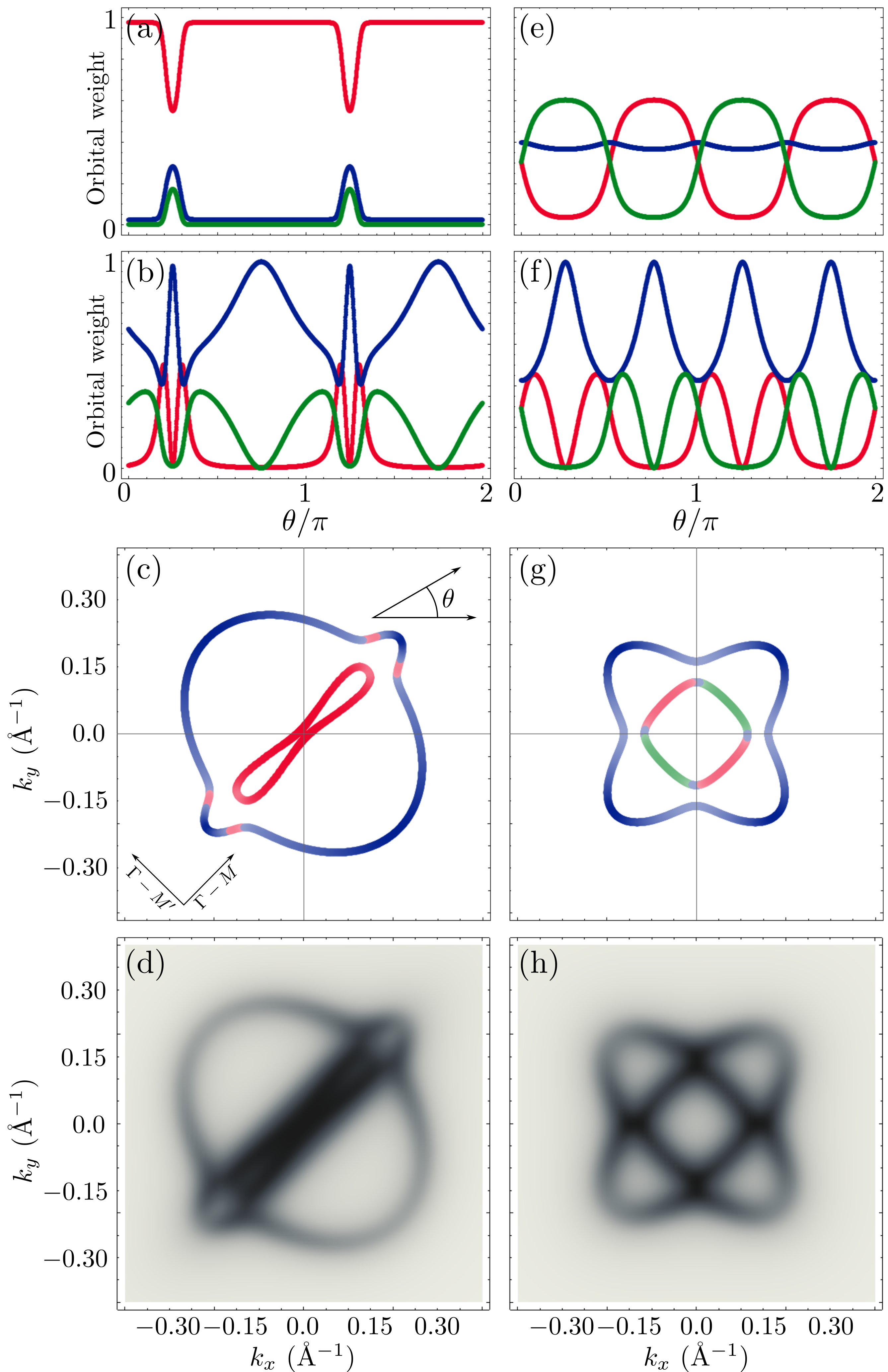} \caption{\label{fig:fs_soc} Fermi surfaces for the SOC scenario, with $\lambda_{{\rm }}=10$~meV.
The orbital color code is the same as Fig. \ref{fig:standard_E}.
(a)--(d) depict the Fermi surface at $T=0$ K, deep in the
nematic phase, while (e)--(h) show the Fermi surface for $T=120$ K, where nematic
order is absent. The top row shows the orbital weight along the Fermi
surfaces, as a function of $\theta$. The lower row shows the coherent
part of the spectral function at the Fermi surface, illustrating the fact that the pocket consisting predominantly of $xy$ fermions is weaker than the one consisting predominantly of $yz$ fermions. In our modeling, this is due to $\Gamma_{xy}>\Gamma_{yz}$.}
\end{figure}

In Fig.~\ref{fig:evolution_at_M_with_T_soc} we show the results of the full calculation of the excitation spectrum for three values of $\phi_3$ (positive, negative, and zero).  The boxes near the lines show the orbital content. We see that one branch of the split upper doublet remains predominantly $yz$ between $T = T_{\mathrm{nem}}$ and $T \ll T_{\mathrm{nem}}$, but the  other becomes predominantly ${xy}$ instead of $xz$. For the lower doublet one branch remains predominantly ${xy}$, but the other becomes $xz$ instead of $xy$. Combining this with the assumption that the $xy$ orbital is incoherent~\cite{lanata13,medici14}, we find that, deep in the nematic state, ARPES should see the sharp coherent $xz$ and $yz$ orbital states, (the latter closer to the Fermi level) and two much weaker incoherent $xy$ excitations in between these two. In Figs. \ref{fig:soc_spectral_function_at_M} and ~\ref{fig:fs_soc} we show the spectral function,
\begin{align}
A(\mbf{k},\omega)=-2\text{Tr}\left[\Im\left(\omega +i\Gamma-\mathcal{H}(\mbf{k})\right)^{-1}\right]\,,
\end{align}
both away and at the Fermi level. Here $\mathcal{H}(\mbf{k})$ is the full Hamiltonian including the nematic order, and $\Gamma$
is a phenomenological diagonal damping term, which we assume to be
larger for the $xy$ orbital, to mimic its incoherence.

The results for the dispersions and the spectral functions in Fig. \ref{fig:soc_spectral_function_at_M} are largely consistent with the available photoemission data. The two stronger peaks at $T \ll T_{\mathrm{nem}}$ have been identified by all photoemission groups~
 ~\cite{watson15,kushnirenko17,watson16,fanfarillo16,fedorov16,rhodes17,
 rhodes18,huh19,yi19}. Ref.~\onlinecite{yi19} identified the orbital content of the excitation closer to the Fermi level as $yz$  and the one farther from the Fermi level as $xz$, by tracking these two excitations between the $M$ and $\Gamma$ points.
Ref.~\onlinecite{fedorov16} reported the observation of two weaker peaks in between the two stronger peaks. In our theory, these are the two $xy$ excitations. As $T$ increases towards $T_{\mathrm{nem}}$, the $yz$ excitation evolves such that it becomes a component of the upper doublet at $T > T_{\mathrm{nem}}$, while the orbital character of the predominantly $xz$ excitation at $T \ll T_{\rm nem}$ evolves as $T$ is increased towards $T_{\rm nem}$ and becomes increasingly $xy$ dominated. Thus, the $yz$ and $xz$ dominated excitations at $T \ll T_{\rm nem}$ do not merge at $T_{\rm nem}$ (see Fig.~\ref{fig:evolution_at_M_with_T_soc}). Instead, the lower component of the upper doublet, which is predominantly $xz$ for $T \geq T_{\rm nem}$ becomes predominantly $xy$ as $T \ll T_{\rm nem}$ and hence becomes increasingly incoherent.
ARPES data did indeed find that near $T_{\mathrm{nem}}$, the excitation closer to the Fermi level is sharper than the one farther from the Fermi level~\cite{coldea18}.

\begin{figure*}
\includegraphics[width=0.95\textwidth]{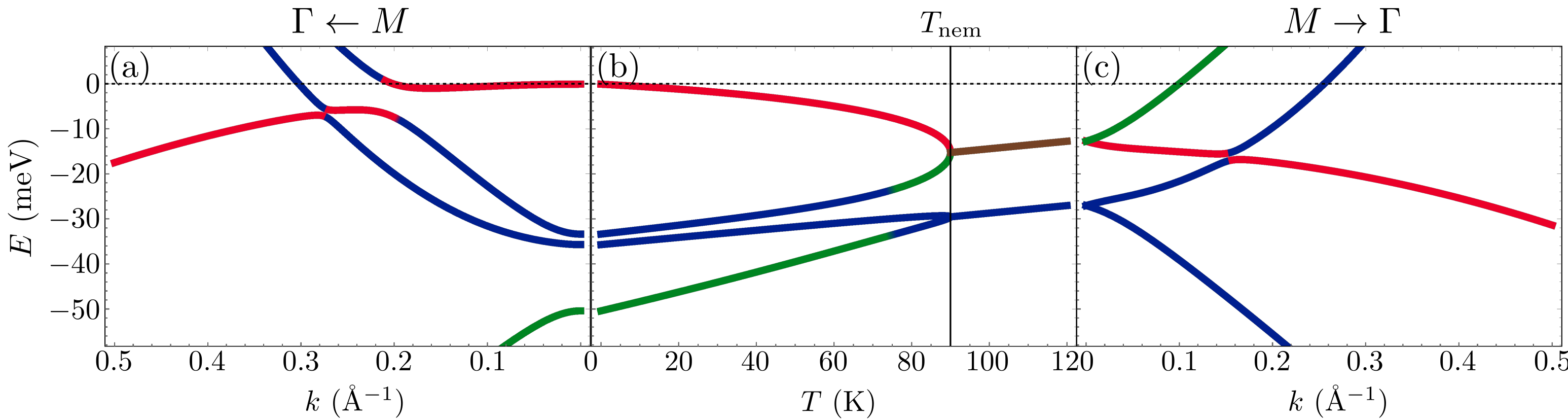}
\caption{Dispersion along $M -\Gamma$ direction at  $T \ll T_{\mathrm{nem}}$ (a) and $T > T_{\mathrm{nem}}$ (c) with the temperature evolution shown in (b). The key effect of nematicity is the drastic change of the orbital composition of the band that crosses the Fermi surface closer to $M$. The orbital color code is the same as in Fig.~\ref{fig:standard_E}.
}
\label{fig:new_1}
\end{figure*}

We also emphasize that the energy splitting of the $xz$ and $yz$ states at low $T$ is not simply $\Delta E =-2\phi_{1}$, as it would be without SOC, but
\begin{eqnarray}
\Delta E &=&
-\phi_1 + \phi_3 + \frac{1}{2} \sqrt{\lambda^2 + (\epsilon_1-\epsilon_3 - \phi_1-\phi_3)^2}
\nonumber \\
&& +\frac{1}{2} \sqrt{\lambda^2 + (\epsilon_1-\epsilon_3 + \phi_1+\phi_3)^2}\,,
\label{eq:E_splitting}
\end{eqnarray}
which is valid in the case $|\phi_1 + \phi_3| > |\epsilon_1 - \epsilon_3|$. The modified splitting is a direct consequence of the orbital transmutation. In Fig.~\ref{fig:fs_soc} we show the two electron Fermi surfaces in the presence of SOC and the orbital composition of the pockets. We see that the smaller, peanut-shaped inner electron pocket is mainly made out of $yz$ fermions, while the outer, more circular-looking pocket is predominantly made out of $xy$ fermions.
The peanut-shaped pocket does contain some admixture
of both $xy$ and $xz$ along the diagonal
direction. The $xz$ contribution is due to SOC. Because of its presence, this portion of the peanut-shaped pocket  becomes visible to ARPES in the polarization orthogonal to $yz$, even if the $xy$ orbital is localized and not detectable. This agrees with Refs.~\onlinecite{watson17,rhodes18}, in which the portions of the peanut-shaped electron-pocket along the diagonal direction have been observed in the polarization orthogonal to $yz$. Note that the degree of $xz$ spectral weight along the diagonal direction is rather sensitive to the magnitudes of the SOC and the nematic order parameters.

In Fig. \ref{fig:new_1}  we show the evolution of the dispersions along the direction from $M$ to $\Gamma$ at low $T \ll T_{\mathrm{nem}}$ and at $T > T_{\mathrm{nem}}$.  We clearly see  that along the $M -\Gamma$ direction, the orbital composition of the band that crosses the Fermi surface closer to $M$ changes drastically between  $T > T_{\mathrm{nem}}$ and $T \ll T_{\mathrm{nem}}$. Orbital transmutation is much weaker along the orthogonal direction, where the main effect of the nematicity is the shrinking of the Fermi momentum for the $yz$ band along the shorter axis of the peanut-shaped Fermi surface.

\section{Excitations in the nematic phase in the presence of surface-induced hybridization}
\label{sec:a2u}

\begin{figure}
\centering \includegraphics[width=0.95\columnwidth]{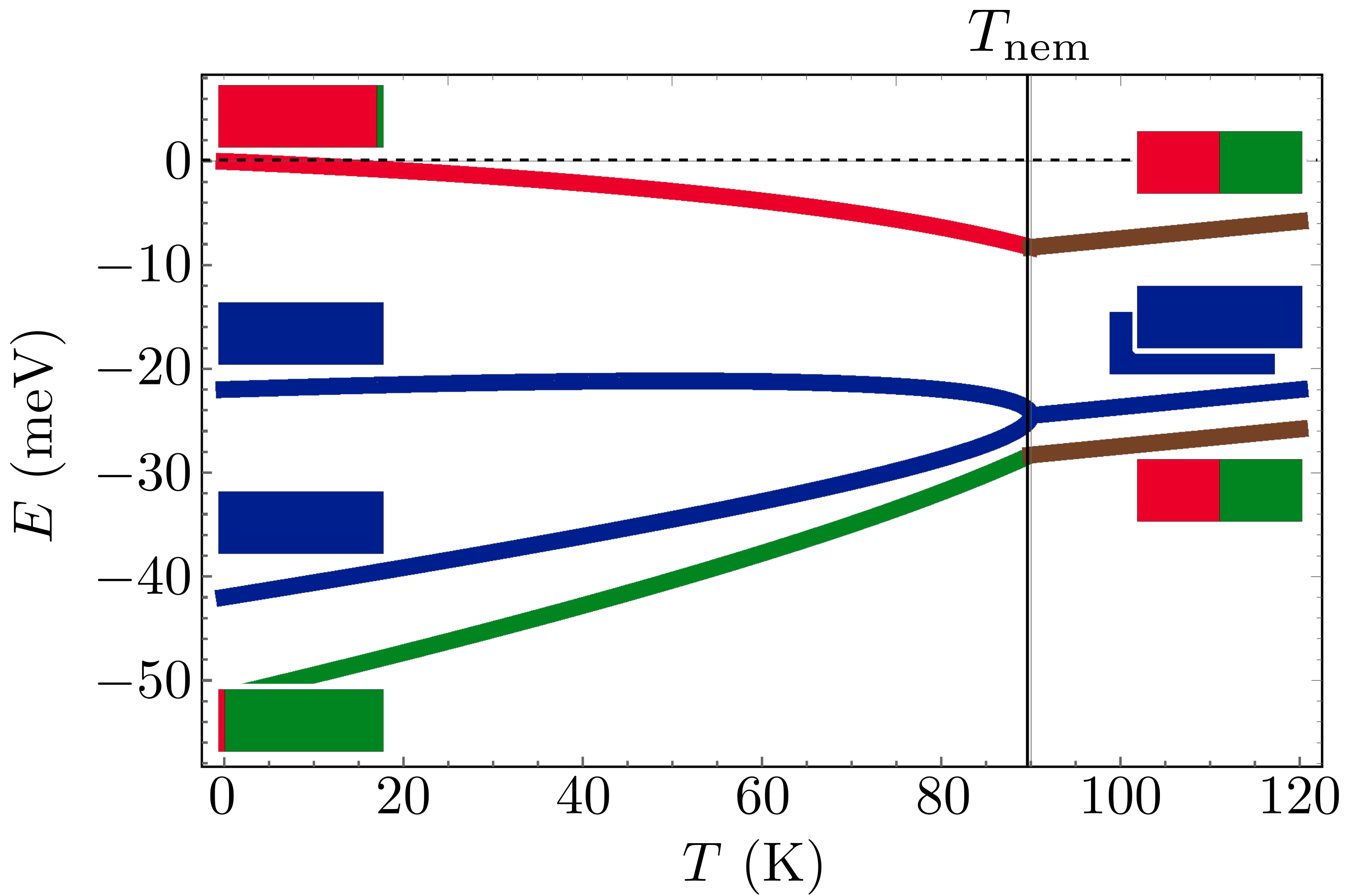}
\caption{\label{fig:evolution_at_M_with_T_a2u} Temperature evolution of the energies at the
$M$-point in the case when the $xz$ and
the $yz$ orbitals hybridize due to surface effects. The orbital color
code is the same as in Fig.~\ref{fig:standard_E}. Here $\epsilon_{1,0}=-26.3$ and $\epsilon_{3,0} = -32.0$ and we chose $\eta = 10$ meV. The doublet with
$xz/yz$ orbital character is split by a surface-induced hybridization
already at temperatures $T>T_{{\rm nem}}$. As nematic order sets in at $T=T_{{\rm nem}}$,
the $yz$ orbital becomes more dominant in the state whose energy
is closest to the Fermi level, while $xz$ is dominant for the state
whose energy is fartherst from the Fermi level, as seen by the difference
between the colored bars at high and low temperatures. Note that the
surface does not break the $xz/yz$ degeneracy above $T_{\mathrm{nem}}$, and the split levels
contain equal weights of the $xz$ and the $yz$ orbitals.}
\end{figure}

In this Section we analyze separately the effect of the surface-induced hybridization (SIH), i.e., of the extra term in the Hamiltonian, given by Eq. (\ref{SIH}). This term, allowed in the surface only, hybridizes fermions in the $xz$ and $yz$ orbitals at $M$ already in the tetragonal phase, where it splits the $xz/yz$ doublet  into two distinct
excitations with mixed $xz+yz$ and $xz-yz$ orbital  character (see Fig.~\ref{fig:evolution_at_M_with_T_a2u}).

Below $T_{{\rm nem}}$, the upper excitation  shifts closer
to the Fermi level, and its orbital content changes from
an equal mixture of $xz$ and $yz$ to almost pure $yz$.
The lower excitation shifts farther from the Fermi level, and its orbital
content changes to almost pure $xz$. At low $T$, the two energies are split by
$\Delta E = 2\sqrt{\phi_{1}^{2}+\eta^{2}}$.  As $T$ increases towards $T_{{\rm nem}}$, the two excitations get closer ro each other, but remain split by
$\Delta E = 2\eta$ also above the nematic transition, at $T>T_{{\rm nem}}$.

\begin{figure}
\centering \includegraphics[width=0.95\columnwidth]{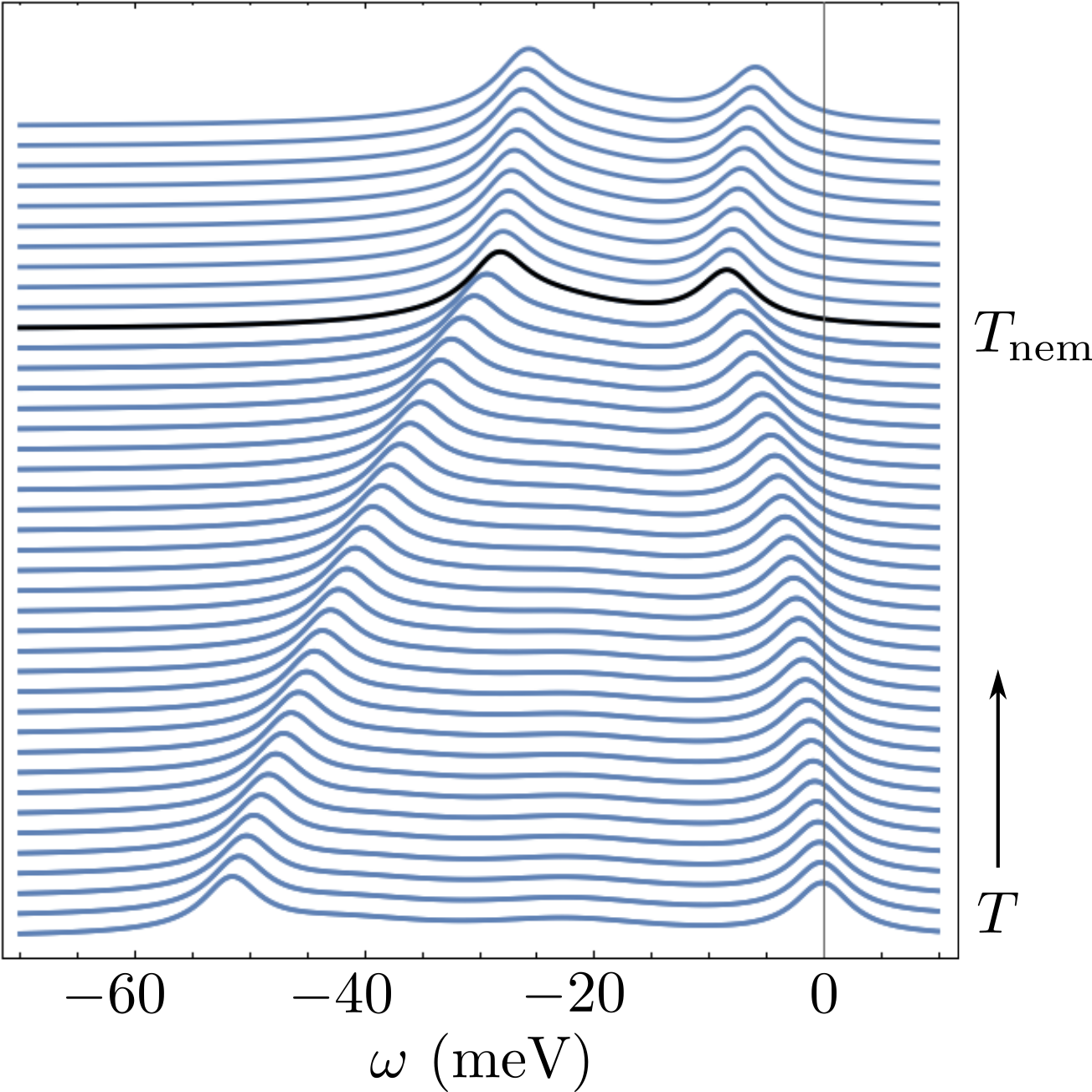} \caption{\label{fig:a2u_spectral_function_at_M} Spectral function as function
of $T$ and $\omega$ in the presence of the surface-induced hybridization
term. The two peaks are clearly distinguished above the nematic
transition and each can be traced all the way down to $T=0$ K. The
two shallow peaks, originating from the $xy$ modes, are at
$\omega\approx-20$ and $\omega\approx-40$ meV at $T=0$.}
\end{figure}

In Fig. \ref{fig:a2u_spectral_function_at_M} we show the spectral function in the presence of SIH. Like before, we set the damping rate to be larger for $xy$ fermions to mimic their incoherence.
The behavior is quite similar to that in the case of SOC. The two differences are (i) the $xz$ excitation remains coherent at all $T$, and (ii)
the $xz$ and $yz$ excitations do not merge even if we extrapolate their positions based only on the low $T$ results.
\begin{figure}
\includegraphics[width=1\columnwidth]{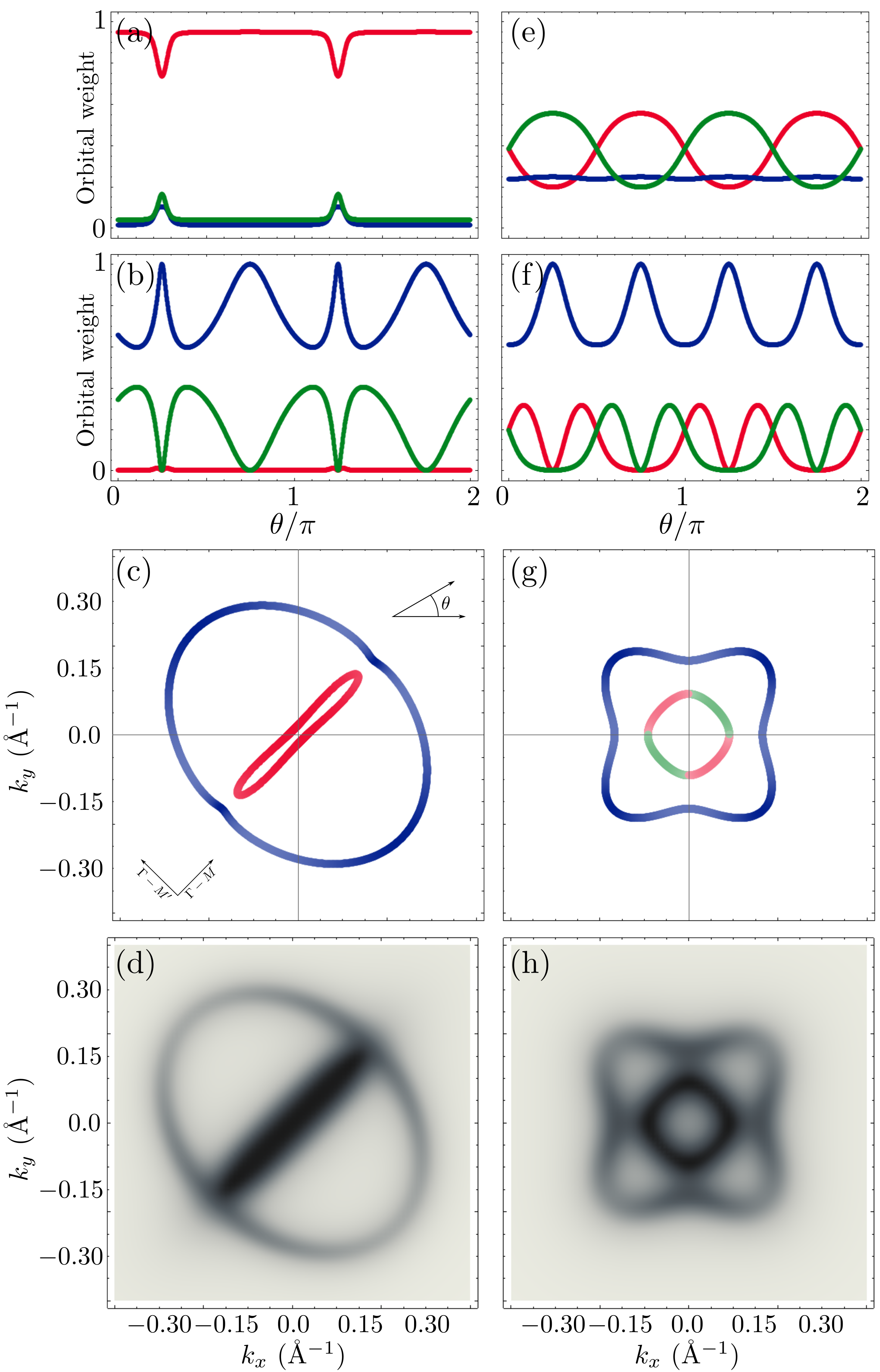} \caption{\label{fig:fs_a2u} Fermi surfaces in the presence of a surface-induced
hybridization term, $\eta=10$~meV. The orbital color code is the
same as Fig. \ref{fig:standard_E}. (a)--(d) correspond to
$T=0$ K, deep in the nematic phase, while (e)--(h) depict the case at $T=120$ K, where nematic order is absent. The Fermi surfaces depicted
here are rather similar to the SOC scenario of Fig. \ref{fig:fs_soc},
although the specific orbital weight around the Fermi pockets differ
slightly.}
\end{figure}
The Fermi surfaces, shown in Fig.~\ref{fig:fs_a2u} along with the orbital weights, is also similar to the case of finite SOC. For temperatures far below $T_{{\rm nem}}$, the inner pocket remains
predominantly of $yz$ character and the outer pocket has predominantly $xy$ character.

\section{Discussion and conclusions}

\label{sec:conclusions}
In this paper we analyzed the puzzling ARPES data on FeSe of the electron pockets at the $M= (\pi,\pi)$ point. Deep in the nematic phase, at $T \ll T_{{\rm nem}}$, ARPES experiments have detected two relatively sharp excitations at $M$~\cite{watson15,watson16,fedorov16,rhodes17,rhodes18,huh19,yi19}, and a recent study unambiguously identified these excitations as having predominantly $xz$ and $yz$ orbital character~\cite{yi19}. Above the nematic transition, at $T> T_{{\rm nem}}$, the $xz$ and $yz$ orbital states are degenerate, and excitations associated with the $xz$ and $yz$ orbitals should therefore merge in the tetragonal phase. However, the data show that the two excitations come closer to each other, but remain split at $T> T_{{\rm nem}}$. We argue that these results are reproduced if we include the effect of SOC.

\begin{figure}
\includegraphics[width=0.95\columnwidth]{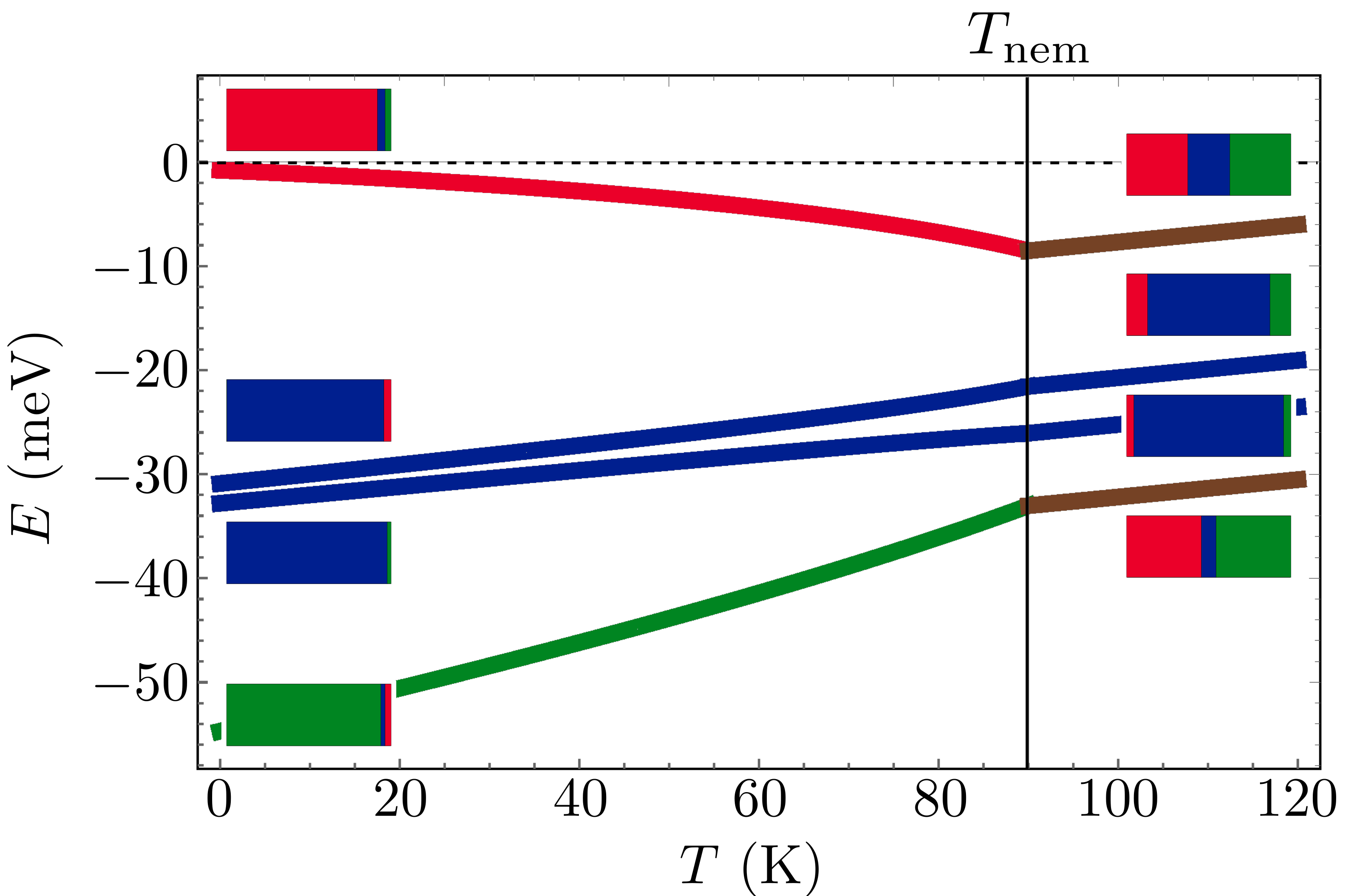}
\caption{\label{fig:evolution_at_M_with_T_soc_a2u} Temperature evolution of the energies at the $M$ point in the presence of both SOC and SIH terms, $\lambda = \eta =10$ meV. Here $\epsilon_{1,0}=-27.6$ meV and $\epsilon_{3,0} = -32.0$ meV. In this case, both doublets are split at $T > T_{\rm nem}$. The $xz/yz$ doublet is split by the hybridization and the $xy$ doublet inherits this splitting through the SOC.}
\end{figure}

\begin{figure}
\includegraphics[width=0.95\columnwidth]{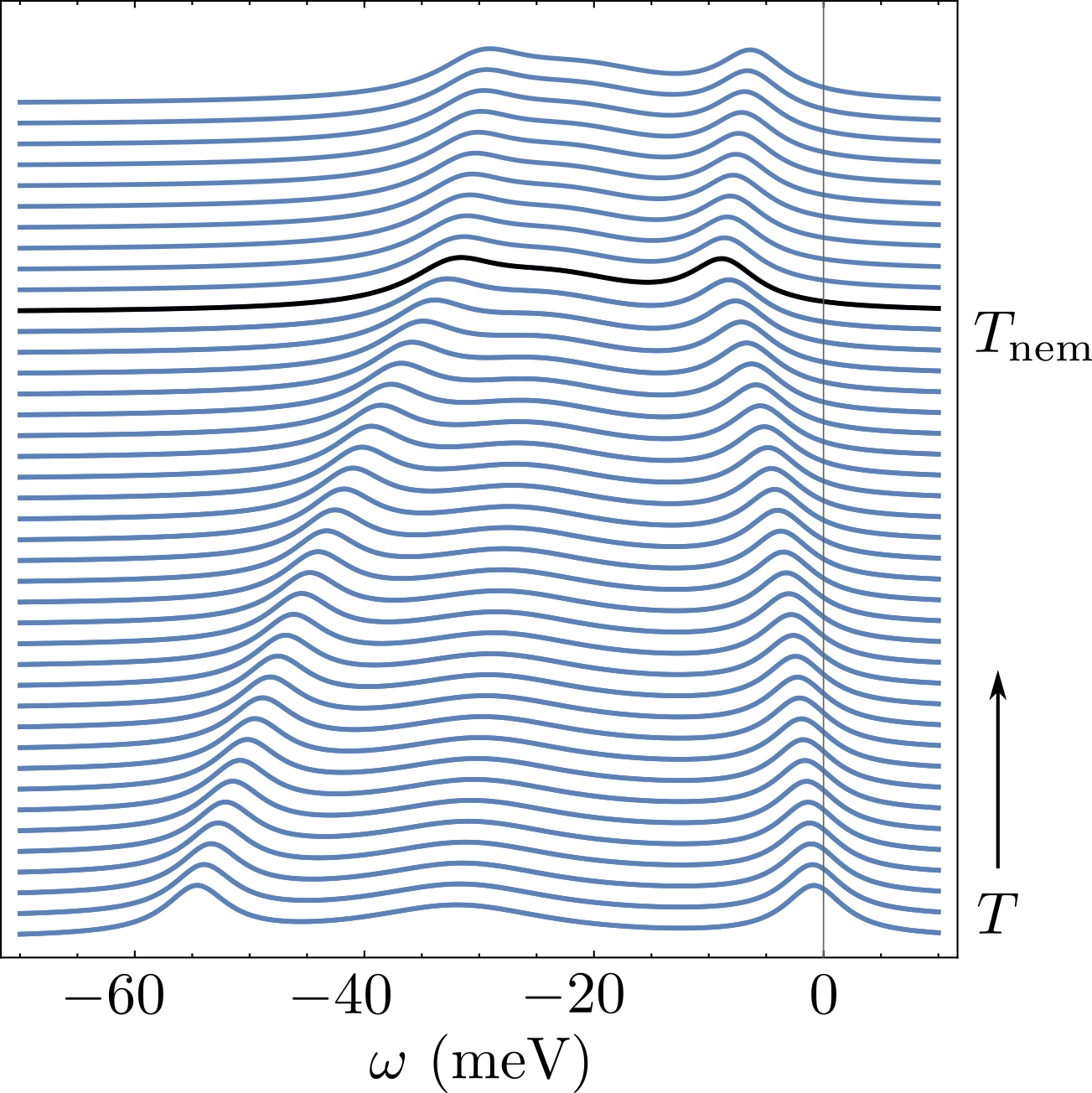}
\caption{\label{fig:soc_a2u_spectral_func_at_M} Spectral function as a function of $T$ and $\omega$ in the presence of both SOC and SIH. As in Fig.~\ref{fig:a2u_spectral_function_at_M}, two peaks can be tracked from $T>T_{\rm nem}$ all the way to $T=0$. Note that the peaks associated with the $xy$-dominated excitations are more visible due to increased orbital weight of the $xz$ and $yz$ components (see Fig.~\ref{fig:evolution_at_M_with_T_soc_a2u}).}
\end{figure}

In the presence of SOC, the excitations at $M$ form two doublets at $T> T_{{\rm nem}}$. The upper doublet is predominantly made out of either the $xz$ or the $yz$ orbital, each with small admixtures of $xy$. The lower doublet consists predominantly of $xy$ fermions, with small admixtures of either $xz$ or $yz$ fermions. Below $T_{{\rm nem}}$, the components of each doublet are split, and there are four distinct excitations. We showed that the dominant orbital character of two of the excitations at $M$ changes drastically between $T > T_{{\rm nem}}$ and $T \ll T_{{\rm nem}}$. Namely, one of the excitations consisting predominantly of the $xy$ orbital along with a small admixture of $xz$ in the tetragonal phase, becomes near-$xz$ deep in the nematic phase. Similarly, the excitation dominated by the $xz$ orbital with a small admixture of $xy$ in the tetragonal phase, becomes near-$xy$ deep in the nematic phase. This phenomenon is due to an effect dubbed orbital transmutation. Because of this effect, the excitation dominated by the $xz$ orbital deep in the nematic phase changes orbital content to $xy$ as $T$ is increased towards $T_{\rm nem}$, and ultimately merges with the lower doublet. As a result, the excitations dominated by the $xz$ and $yz$ orbitals at $T \ll T_{\rm nem}$ do not merge at $T_{\rm nem}$, in agreement with ARPES data. Other features of the spectral function at $M$, shown in Fig.~\ref{fig:soc_spectral_function_at_M}, and the shape and orbital composition of the two electron pockets, shown in Fig.~\ref{fig:fs_soc}, also agree with the data. We showed that almost the same behavior is obtained if, instead of SOC, we include the SIH between the $xz$ and $yz$ orbitals at $M$. Such hybridization is not allowed in the bulk because it would violate the glide-plane symmetry, but is allowed at the surface. The only substantial difference between the effects of SOC and SIH is that, in the presence of SOC, there are two doublets at $T>T_{{\rm nem}}$, while in the presence of SIH there is one $xy$ doublet~\cite{footnote_sih} and two singlets with equal superpositions of $xz$ and $yz$. For completeness, in Fig.~\ref{fig:evolution_at_M_with_T_soc_a2u} we illustrate the combined effect of SOC and SIH on the excitation spectrum, and in Fig.~\ref{fig:soc_a2u_spectral_func_at_M} we plot the spectral function for this scenario. Whether the observed behavior is due to SOC or SIH or a combination of both requires further experiments probing the orbital content of the excitations at $T>T_{{\rm nem}}$.
We note in this regard that the combination of SOC and SIH  breaks double degeneracy of the bands away from the $M$ point. This results in the doubling of the number of Fermi surfaces. The latter has not been  resolved in experiments, which likely implies that the combined effect of SOC and SIH is small and is masked by the thermal broadening. We also note that at the $\Gamma$ point, both Rashba- and Dresselhaus-like SOC terms are generated for momenta away from the $\Gamma$ point, as discussed in Ref.~\onlinecite{christensen19}. These terms are responsible for lifting the double degeneracy of the bands in the vicinity of $\Gamma$ and for  doubling of the number of hole Fermi surfaces.

Neither SOC on the electron pockets, $\lambda$, nor the SIH, $\eta$, has been measured directly. For the hole pockets, SOC has been measured and is around $20$~meV~\cite{borisenko16}. We showed the results for $\lambda=10$ meV. As $\lambda$ increases, the orbital transmutation becomes more effective, but qualitatively the results do not change.
The same holds if we only include SIH -- our results are shown for $\eta  = 10$ meV, but variation of $\eta$ does not lead to qualitative changes. We also argued that the larger electron pocket (the one with non-peanut-like shape) is not seen in ARPES experiments because $xy$ fermions are incoherent. 

The two scenarios can be distinguished by comparing measurements at the $M$ point in the tetragonal phase with measurements in the orthorhombic phase. In the SOC scenario, there are two doubly degenerate bands in the tetragonal phase, while in the SIH scenario there are four bands with different energies, compare Figs. \ref{fig:evolution_at_M_with_T_soc} and \ref{fig:evolution_at_M_with_T_soc_a2u} (recall that Fig.~\ref{fig:evolution_at_M_with_T_a2u} assumes a vanishing SOC). Additionally, in the SOC scenario, one doubly degenerate band consists predominantly of the $xz$ and $yz$ orbitals, while the other is mostly $xy$. If the $xy$ orbital is more incoherent, the ARPES intensity of the $xz/yz$ mode should be higher than the intentisty of the $xy$ mode, and the width of the first mode should be narrower. In the SIH scenario, two modes in the tetragonal phase are made equally of $xz$ and $yz$ orbitals, and their intensity should be equal, even if the other two modes (with $xy$ character) are too incoherent to be detected. If both effects are present and are of comparable strength, ARPES measurements would reveal four peaks of varying intensity, depending on their relative orbital content (see Fig.~\ref{fig:evolution_at_M_with_T_soc_a2u}).

A different reason for the disappearance of the second electron-pocket was put forward in Ref.~\onlinecite{yi19}. There, it was argued that  it shrinks in the nematic phase. This is in variation with our result that the size of this pocket increases below $T_{{\rm nem}}$. The reduction of the non-peanut-shaped pocket can in principle be obtained if we assume that the magnitude of $\phi_3$ is comparable to $\epsilon_3$. However, this requires fine-tuning, and this scenario does not explain the non-merging of the $xz$ and $yz$ excitations at $T_{\rm nem}$.

Another seemingly appealing option is to introduce SOC or hybridization between $xz/xy$ and $yz/xy$ fermions within the same pocket in the 1-Fe BZ , i.e. add to the Hamiltonian the terms
\begin{eqnarray}
\mathcal{H}_{{\rm intra-pocket\ hyb}} & = & \eta_{{\rm hyb}}^{(1)}\hat{d}_{xz,\sigma}^{\dagger}\hat{d}_{xy,\sigma}\nonumber \\
 & + & \eta_{{\rm hyb}}^{(2)}d_{yz,\sigma}^{\dagger}d_{xy,\sigma}+\text{H.c.}\,,\label{eq:intra_poc_hyb}\\
\mathcal{H}_{{\rm intra-pocket\ SOC}} & = & \lambda_{{\rm SOC}}^{(1,i)}\hat{d}_{xz,\alpha}^{\dagger}\sigma_{\alpha\beta}^{i}\hat{d}_{xy,\beta}\nonumber \\
 & + & \lambda_{{\rm SOC}}^{(2,i)}d_{yz,\alpha}^{\dagger}\sigma_{\alpha\beta}^{i}d_{xy,\beta}+\text{H.c.}\,.\label{eq:intra_poc_soc}
\end{eqnarray}
This would split the $xz/yz$ and $xy$ doublets and shrink the size
of the non-peanut-shape pocket~\cite{yi19,huh19}. However,
we emphasize that neither Eq.~\eqref{eq:intra_poc_hyb} nor Eq.~\eqref{eq:intra_poc_soc} are allowed by the $P4/nmm$ space group symmetry, regardless of whether the system is in the tetragonal or in the nematic phase~\cite{cvetkovic13}. At the surface, $\mathcal{H}_{{\rm intra-pocket\ hyb}}$ is still not allowed, but $\mathcal{H}_{{\rm intra-pocket\ SOC}}$ with $i = x,y$ is allowed~\cite{christensen19}. However, this term is generated by the combination of atomic SOC, Eq.~\eqref{eq:soc}, and SIH, Eq.~\eqref{SIH} and the effects of this term are secondary to those from the SOC and the SIH (see Figs.~\ref{fig:evolution_at_M_with_T_soc_a2u} and \ref{fig:soc_a2u_spectral_func_at_M}).

\begin{acknowledgments}
We are grateful to Brian Andersen, Lara Benfatto, Anna B{\"o}hmer, Sergey Borisenko,
Peter Hirschfeld, Jian Kang, Timur Kim, Andreas Kreisel, Dung-Hai
Lee, Luke Rhodes, Matthew Watson, and Ming Yi for stimulating discussions. M.H.C.
and R.M.F. were supported by the U.S. Department of Energy, Office
of Science, Basic Energy Sciences, under Award No. DE-SC0012336. A.V.C.
is supported by U.S. Department of Energy, Office of Science, Basic
Energy Sciences, under Award No. DE-SC0014402.
\end{acknowledgments}


\begin{thebibliography}{1}

\bibitem{bohmer17} A. E. B{\"o}hmer and A. Kreisel. Nematicity, magnetism and superconductivity in FeSe. J. Phys. Cond. Mat. {\bf 30}, 023001 (2017).

\bibitem{sprau17} P. O. Sprau, A. Kostin, A. Kreisel, A. E. B{\"o}hmer, V. Taufour, P. C. Canfield, S. Mukherjee, P. J. Hirschfeld, B. M. Andersen, and J. C. S{\'e}amus Davis. Discovery of orbital-selective Cooper pairing in FeSe. Science {\bf 357}, 75 (2017).

\bibitem{coldea18} A. I. Coldea and M. D. Watson. The Key Ingredients of the Electronic Structure of FeSe. Ann. Rev. Cond. Mat. Phys. {\bf 9}, 125 (2018).

\bibitem{sun2016} J. P. Sun, \textit{et al.} Dome-shaped magnetic order competing with high-temperature superconductivity at high pressures in FeSe. Nat. Commun. {\bf 7}, 12146 (2016).

\bibitem{bohmer2018} A. E. B{\"o}hmer, \textit{et al.} Distinct pressure evolution of coupled nematic and magnetic order in FeSe. arXiv:1803.09449, (2018).

\bibitem{mcqueen09} T. M. McQueen, A. J. Williams, P. W. Stephens, J. Tao, Y. Zhu, V. Ksenofontov, F. Casper, C. Felser, and R. J. Cava. Tetragonal-to-Orthorhombic Structural Phase Transition at 90 K in the Superconductor Fe$_{1.01}$Se. Phys. Rev. Lett. {\bf 103}, 057002 (2009).

\bibitem{hsu08} F.-C Hsu, J.-Y. Luo, K.-W. Yeh, T.-K. Chen, T.-W. Huang, P. M. Wu, Y.-C. Lee, Y.-L. Huang, Y.-Y. Chu, D.-C. Yan, and M.-K. Wu. Superconductivity in the PbO-type structure $\alpha$-FeSe. Proc. Natl. Acad. Sci. {\bf 105}, 14262 (2009).

\bibitem{footnote_notation} Strictly speaking, the notations $\Gamma$, X and Y hold only for $k_{z}=0$. To simplify the presentation we will use this notation without specifying the value of $k_{z}$.

\bibitem{lin11} C.-H. Lin, T. Berlijn, L. Wang, C.-C. Lee, W.-G. Yin, and W. Ku. One-Fe versus Two-Fe Brillouin Zone of Fe-Based Superconductors: Creation of the Electron Pockets by Translational Symmetry Breaking. Phys. Rev. Lett. {\bf 107}, 257001 (2011).

\bibitem{cvetkovic13} V. Cvetkovic and O. Vafek. Space group symmetry,
spin-orbit coupling, and the low-energy effective Hamiltonian for
iron-based superconductors. Phys. Rev. B \textbf{88}, 134510 (2013).

\bibitem{christensen15} M. H. Christensen, J. Kang, B. M. Andersen, I. Eremin, and R. M. Fernandes. Spin reorientation driven by the interplay between spin-orbit coupling and Hund's coupling in iron pnictides. Phys. Rev. B {\bf 92}, 214509 (2015).

\bibitem{borisenko16} S. V. Borisenko, D. V. Evtushinsky, Z.-H. Liu,
I. Morozov, R. Kappenberger, S. Wurmehl, B. B{ü}chner, A. N. Yaresko,
T. K. Kim, M. Hoesch, T. Wolf, and N. D. Zhigadlo. Direct observation
of spin-orbit coupling in iron-based superconductors. Nat. Phys. \textbf{12}, 311 (2016).

\bibitem{terashima14} T. Terashima, \textit{et al.} Anomalous Fermi surface in FeSe seen by Shubnikov-de Haas oscillation measurements. Phys. Rev. B {\bf 90}, 144517 (2014).

\bibitem{audouard15} A. Audouard, F. Duc, L. Drigo, P. Toulemonde, S. Karlsson, P. Strobel, and A. Sulpice. Quantum oscillations and upper critical magnetic field of the iron-based superconductor FeSe. Europhys. Lett. {\bf 109}, 27003 (2015).

\bibitem{maletz14} J. Maletz, \textit{et al.} Unusual band renormalization in the simplest iron-based superconductor FeSe$_{1-x}$. Phys. Rev. B {\bf 89}, 220506(R) (2014).

\bibitem{rhodes17} L. C. Rhodes, M. D. Watson, A. A. Haghighirad, M. Eschrig, and T. K. Kim. Strongly enhanced temperature dependence of the chemical potential in FeSe. Phys. Rev. B {\bf 95}, 195111 (2017).

\bibitem{kushnirenko17} Y. S. Kushnirenko, A. A. Kordyuk, A. V. Fedorov, E. Haubold, T. Wolf, B. B{\"u}chner, and S. V. Borisenko. Anomalous temperature evolution of the electronic structure of FeSe. Phys. Rev. B {\bf 96}, 100504(R) (2017).

\bibitem{watson15} M. D. Watson, T. K. Kim, A. A. Haghighirad, N. R. Davies, A. McCollam, A. Narayanan, S. F. Blake, Y. L. Chen, S. Ghannadzadeh, A. J. Schofield, M. Hoesch, C. Meingast, T. Wolf, and A. I. Coldea. Emergence of the nematic electronic state in FeSe. Phys. Rev. B {\bf 91}, 155106 (2015).

\bibitem{shimojima14} T. Shimojima, \textit{et al.} Lifting of $xz/yz$ orbital degeneracy at the structural transition in detwinned FeSe. Phys. Rev. B {\bf 90}, 121111(R) (2014).

\bibitem{christensen19} M. H. Christensen, J. Kang, and R. M. Fernandes.
Intertwined spin-orbital coupled orders in the iron-based superconductors. Phys. Rev. B {\bf 100}, 014512 (2019).

\bibitem{suzuki15} Y. Suzuki, \textit{et al.} Momentum-dependent sign inversion of orbital order in superconducting FeSe. Phys. Rev. B {\bf 92}, 205117 (2015).

\bibitem{fanfarillo16} L. Fanfarillo, J. Mansart, P. Toulemonde, H. Cercellier, P. Le F{\`e}vre, F. Bertran, B. Valenzuela, L. Benfatto, and V. Brouet. Orbital-dependent Fermi surface shrinking as a fingerprint of nematicity in FeSe. Phys. Rev. B {\bf 94}, 155138 (2016).

\bibitem{fedorov16} A. Fedorov, A. Yareshko, T. K. Kim, Y. Kushnirenko,
E. Haubold, T. Wolf, M. Hoesch, A. Gr{ü}neis, B. B{ü}chner, and
S. V. Borisenko. Effect of nematic ordering on electronic structure
of FeSe. Sci. Rep. \textbf{6}, 36834 (2016).

\bibitem{watson17} M. D. Watson, A. A. Haghighirad, L. C. Rhodes, M. Hoesch, and T. K. Kim. Electronic anisotropies revealed by detwinned angle-resolved photo-emission spectroscopy measurements of FeSe. New J. Phys. {\bf 19}, 103021 (2017).

\bibitem{watson16} M. D. Watson, T. K. Kim, L. C. Rhodes, M. Eschrig, M. Hoesch, A. A. Haghighirad, and A. I. Coldea. Evidence for unidirectional nematic bond ordering in FeSe. Phys. Rev. B {\bf 94}, 201107(R) (2016).

\bibitem{yi19} M. Yi, \textit{et al.} The Nematic Energy Scale and the Missing Electron Pocket in FeSe. arXiv:1903.04557 (2019).

\bibitem{huh19} S. Huh, J. Seo, B. Kim, S. Cho, J. Jung, S. Kim, Y. Koh, C. Kwon, J. Kim, W. Kyung, J. D. Denlinger, Y. Kim, B. Chae, N. Kim, Y. Kim, C. Kim. Lifted electron pocket and reversed orbital occupancy imbalance in FeSe. arXiv:1903.08360 (2019).

\bibitem{vafek14} O. Vafek and R. M. Fernandes. Distinguishing spin-orbit coupling and nematic order in the electronic spectrum of iron-based superconductors. Phys. Rev. B {\bf 90}, 214514 (2014).

\bibitem{kang18} J. Kang, R. M. Fernandes, and A. V. Chubukov. Superconductivity in FeSe: The Role of Nematic Order. Phys. Rev. Lett. {\bf 120}, 267001 (2018).

\bibitem{kontani_1} S. Onari, Y. Yamakawa, and H. Kontani. Sign-Reversing Orbital Polarization in the Nematic Phase of FeSe due to the C$_2$ Symmetry Breaking in the Self-Energy.
Phys. Rev. Lett. {\bf 116}, 227001 (2016).

\bibitem{latest} H. Pfau, S. D. Chen, M. Yi, M. Hashimoto, C. R. Rotundu, J. C. Palmstrom, T. Chen, P.-C. Dai, J. Straquadine, A. Hristov, R. J. Birgeneau, I. R. Fisher, D. Lu, and Z.-X. Shen. Momentum Dependence of the Nematic Order Parameter in Iron-Based Superconductors. Phys. Rev. Lett. {\bf 123}, 066402 (2019).

\bibitem{chubukov16} A. V. Chubukov, M. Khodas, and R. M. Fernandes. Magnetism, Superconductivity, and Spontaneous Orbital Order in Iron-Based Superconductors: Which Comes First and Why? Phys. Rev. X {\bf 6}, 041045 (2016).

\bibitem{xing16} R.-Q. Xing, L. Classen, M. Khodas, and A. V. Chubukov. Competing instabilities, orbital ordering, and splitting of band degeneracies from a parquet renormalization group analysis of a four-pocket model for iron-based superconductors: Application to FeSe.
Phys. Rev. {\bf 95}, 085108 (2017).

\bibitem{rhodes18} L. C. Rhodes, M. D. Watson, A. A. Haghighirad, D. V. Evtushinsky, M. Eschrig, and T. K. Kim. Scaling of the superconducting gap with orbital character in FeSe. Phys. Rev. B {\bf 98}, 180503(R) (2018).

\bibitem{footnote_sih} The SIH term also splits the two $xy$ modes above the nematic transition~\cite{christensen19}. For simplicity we ignore this effect here.

\bibitem{kreisel17} A. Kreisel, B. M. Andersen, P. O. Sprau, A. Kostin, J. C. S{\'e}amus Davis, and P. J. Hirschfeld. Orbital selective pairing and gap structures of iron-based superconductors. Phys. Rev. B {\bf 95}, 174504 (2017).

\bibitem{benfatto18} L. Benfatto, B. Valenzuela, and L. Fanfarillo. Nematic pairing from orbital-selective spin fluctuations in FeSe. npj Quant. Mat. {\bf 3}, 56 (2018).

\bibitem{lanata13} N. Lanat{\`a}, H. U. R. Strand, G. Giovannetti, B. Hellsing, L de' Medici, and M. Capone. Orbital selectivity in Hund's metals: The iron chalcogenides. Phys. Rev. B {\bf 87}, 045122 (2013).

\bibitem{medici14} L. de' Medici, G. Giovannetti, and M. Capone. Selective Mott Physics as a Key to Iron Superconductors. Phys. Rev. Lett. {\bf 112}, 177001 (2014).

\bibitem{yu17} R. Yu and Q. Si. Orbital-selective Mott phase in multiorbital models for iron pnictides and chalcogenides. Phys. Rev. B {\bf 96}, 125110 (2017).

\bibitem{kushnirenko18} Y. S. Kushnirenko, A. V. Fedorov, E. Haubold, S. Thirupathaiah, T. Wolf, S. Aswartham, I. Morozov, T. K. Kim, B. B{\"u}chner, and S. V. Borisenko. Three-dimensional superconducting gap in FeSe from angle-resolved photoemission spectroscopy. Phys. Rev. B {\bf 97}, 180501(R) (2018).

\bibitem{eugenio18} P. M. Eugenio and O. Vafek. Classification of symmetry derived pairing at the $M$ point in FeSe. Phys. Rev. B {\bf 98}, 014503 (2018).

\bibitem{Kang16} J. Kang and R. M. Fernandes. Superconductivity in FeSe Thin Films Driven by the Interplay between Nematic Fluctuations and Spin-Orbit Coupling.  Phys. Rev. Lett {\bf 117}, 217003 (2016).

\bibitem{Agterberg17} J. O'Halloran, D. F. Agterberg, M. X. Chen, and M. Weinert. Stabilizing the spin vortex crystal phase in two-dimensional iron-based superconductors. Phys. Rev. B {\bf 95}, 075104 (2017).

\bibitem{footnote_other_domain} In the other domain, the situation is reversed, as $\phi_{\Gamma}<0$ and $\phi_1 >0$. In this case, the $yz$ dominated excitation undergoes an orbital transmutation.

%

\end{thebibliography}
\end{document}